\documentclass[twocolumn,preprintnumbers,a4paper,
superscriptaddress,floatfix,twoside]{revtex4}

\usepackage{bm}
\usepackage{epsfig}
\usepackage{graphics}
\usepackage{natbib}

\usepackage{fancyh}
\usepackage[l]{floatflt}
\usepackage{epsfig}
\usepackage{amssymb}
\usepackage{latexsym}
\usepackage{times}
\usepackage{slashed}
\usepackage{upgreek}
\usepackage{amsmath}
\usepackage{amsfonts}
\usepackage{amsbsy}
\usepackage{amscd}
\usepackage{bbm}

\newcommand{\eec}{\end{center}}
\newcommand{\bec}{\begin{center}}

\newcommand{\eem}{\end{matrix}}
\newcommand{\bem}{\begin{matrix}}
\newcommand{\eeq}{\end{equation}}
\newcommand{\beq}{\begin{equation}}
\newcommand{\ba}{\begin{array}}
\newcommand{\ea}{\end{array}}
\newcommand{\bea}{\begin{eqnarray}}
\newcommand{\eea}{\end{eqnarray}}
\newcommand{\baq}{\begin{eqnarray}}
\newcommand{\eaq}{\end{eqnarray}}
\newcommand{\beqs}{\begin{subequations}}
\newcommand{\eeqs}{\end{subequations}}

\newcommand\eqs[2]{Eqs.~(\ref{#1}) and (\ref{#2})}

\newcommand\eqss[3]{Eqs.~(\ref{#1}), (\ref{#2}) and (\ref{#3})}

\newcommand{\ftn}{\footnotesize}

\newcommand{\ssz}{\scriptsize}
\newcommand{\TeV}{{\mbox{\rm TeV}}}

\newcommand{\GeV}{{\mbox{\rm GeV}}}
\newcommand{\eV}{{\mbox{\rm eV}}}
\newcommand{\sFref}[2]{Fig.~\ref{#1}-{\small \sf ({#2})}}
\newcommand{\sEref}[2]{Eq.~(\ref{#1}{\small\sf {#2}})}

\newcommand{\etal}{{\it et al.\/}}

\def\dag{\dagger}
\def\llgm{\left\lgroup}
\def\rrgm{\right\rgroup}
\def\lf{\left(}
\def\rg{\right)}
\newcommand\vev[1]{\langle {#1} \rangle}
\newcommand{\Gr}{\ensuremath{\widetilde{G}}}

\newcommand{\Vhi}{\ensuremath{\widehat V_{\rm CI}}}
\newcommand{\Hhi}{\ensuremath{\widehat H_{\rm CI}}}

\newcommand{\Ohi}{\ensuremath{\Omega_{\rm CI}}}
\newcommand{\Khi}{\ensuremath{K}}

\newcommand{\Vhio}{\ensuremath{\widehat V_{\rm CI0}}}

\newcommand{\mP}{\ensuremath{m_{\rm P}}}
\newcommand{\Mpq}{\ensuremath{M_{\rm PQ}}}
\newcommand{\la}{\ensuremath{\lambda_a}}
\newcommand{\lm}{\ensuremath{\lambda_\mu}}

\newcommand{\ck}{\ensuremath{c_\mathcal{R}}}
\newcommand{\kx}{\ensuremath{k_{\bar P}}}

\newcommand{\mpq}{\ensuremath{m_{\rm PQ}}}

\newcommand{\Gsn}{\ensuremath{\Gamma_{\rm I}}}
\newcommand{\msn}{\ensuremath{m_{\rm I}}}

\newcommand{\hd}{{\ensuremath{H_d}}}
\newcommand{\hu}{{\ensuremath{H_u}}}

\newcommand{\ns}{\ensuremath{n_{\rm s}}}
\newcommand{\as}{\ensuremath{\alpha_{\rm s}}}

\newcommand{\rcc}{\ensuremath{\mathcal{R}}}

\newcommand{\Ve}{\ensuremath{\widehat{V}}}
\newcommand{\He}{\ensuremath{\widehat{H}}}
\newcommand{\Ne}{\ensuremath{\widehat{N}}}
\newcommand{\sni}{\ensuremath{\nu^c_i}}
\newcommand{\sn}{\ensuremath{\tilde \nu^c}}
\newcommand{\ssni}{\ensuremath{\tilde \nu_i^c}}
\newcommand{\what}{\ensuremath{\widehat}}

\newcommand{\dma}{\ensuremath{D_{k{\rm a}}}}
\newcommand{\dbma}{\ensuremath{\bar D_{k{\rm a}}}}
\newcommand{\hla}{\ensuremath{h_{l{\rm a}}}}
\newcommand{\hbla}{\ensuremath{\bar h_{l{\rm a}}}}
\newcommand{\da}{\ensuremath{D_{{\rm a}}}}
\newcommand{\dba}{\ensuremath{\bar D_{{\rm a}}}}
\newcommand{\ha}{\ensuremath{h_{{\rm a}}}}
\newcommand{\hba}{\ensuremath{\bar h_{{\rm a}}}}

\def\ve{\varepsilon}

\def\bbet{{\bar\beta}}
\def\al{{\alpha}}

\def\n{\bar{n}}

\def\th{{\theta}}
\def\bP{{\bar P}}
\def\bX{{\bar \Phi}}
\def\Xb{{\Phi}}

\newcommand{\diag}{\ensuremath{{\sf diag}}}
\newcommand{\im}{\ensuremath{{\sf Im}}}
\newcommand{\br}{\ensuremath{{\sf Br}}}
\newcommand{\tr}{{\mbox{\sf\ssz T}}}

\newcommand{\Trh}{\ensuremath{T_{\rm rh}}}
\newcommand{\sg}{\ensuremath{\sigma}}

\newcommand{\sgf}{\ensuremath{\sigma_{\rm f}}}
\newcommand{\xsg}{\ensuremath{x_{\sigma}}}

\newcommand{\ld}{\ensuremath{\lambda}}
\newcommand{\Ld}{\ensuremath{\Lambda}}

\newcommand{\se}{\ensuremath{\widehat\sigma}}

\newcommand{\hepth}[1]{{\ftn \tt hep-th/#1}}
\newcommand{\hepph}[1]{{\ftn\tt hep-ph/#1}}

\newcommand{\astroph}[1]{{\ftn\tt astro-ph/#1}}
\newcommand{\arxiv}[1]{{\ftn\tt  arXiv:#1}}
\renewcommand{\theparagraph}{{\sf\small\arabic{paragraph}}}
\renewcommand{\thesubsubsection}{{\small\sf\arabic{subsubsection}}}

\newcommand{\Eref}[1]{Eq.~(\ref{#1})}
\newcommand{\Sref}[1]{Sec.~\ref{#1}}
\newcommand{\Fref}[1]{Fig.~\ref{#1}}
\newcommand{\Tref}[1]{Table~\ref{#1}}
\newcommand{\cref}[1]{Ref.~\cite{#1}}

\def\FHI{nMCI~}
\def\Ka{K\"{a}hler potential}

\newcommand\mtn[9]{\ensuremath{\llgm\bem #1&#2&#3\cr #4&#5&#6 \cr #7&#8&#9\eem\rrgm}}

\newcommand{\bdhh}{{\ensuremath{\normalsize I{\kern-2.9pt H}}}}

\newcommand{\mrh[1]}{\ensuremath{M_{#1\nu^c}}}

\newcommand{\mD[1]}{\ensuremath{m_{#1\rm D}}}
\newcommand{\mn[1]}{\ensuremath{m_{#1\nu}}}
\newcommand{\Gm[1]}{\ensuremath{\Gamma_{#1}}}
\newcommand{\sft}{{\rm soft}}

\renewcommand{\uppercase}{\normalsize\bf\scshape}
\renewcommand{\acknowledgmentsname}{\bf\scshape Acknowledgments}
\renewcommand{\figurename}{\bf\scshape Fig.}
\renewcommand{\tablename}{\bf\scshape Table}
\renewcommand{\refname}{{\bf\scshape References}}

\renewenvironment{subequations}{%
\refstepcounter{equation}%
\setcounter{parentequation}{\value{equation}}%
  \setcounter{equation}{0}
  \def\theequation{\theparentequation{\sf\small\alph{equation}}}%
  \ignorespaces
}{%
  \setcounter{equation}{\value{parentequation}}%
  \ignorespacesafterend
}

\begin{document}


\title{\bf\scshape Non-Minimal Chaotic Inflation, Peccei-Quinn
Phase Transition \\ and non-Thermal Leptogenesis}

\author{\scshape Constantinos Pallis}
\affiliation{Department of Physics, University of Cyprus, P.O. Box
20537, Nicosia 1678, CYPRUS
\\  {\sl e-mail address: }{\ftn\tt cpallis@ucy.ac.cy}}
\author{\scshape  Qaisar Shafi} 
\affiliation{ Bartol Research Institute, Department of Physics and
Astronomy, University of Delaware, Newark, DE 19716, USA\\  {\sl
e-mail address: }{\ftn\tt shafi@bartol.udel.edu}}


\begin{abstract}

\noindent {\ftn \bf\scshape Abstract:} We consider a
phenomenological extension of the minimal supersymmetric standard
model (MSSM) which incorporates non-minimal chaotic inflation,
driven by a quadratic potential in conjunction with a linear term
in the frame function. Inflation is followed by a Peccei-Quinn
phase transition, based on renormalizable superpotential terms,
which resolves the strong CP and $\mu$ problems of MSSM and
provide masses lower than about $10^{12}~\GeV$ for the
right-handed (RH) (s)neutrinos. Baryogenesis occurs via
non-thermal leptogenesis, realized by the out-of-equilibrium decay
of the RH sneutrinos, which are produced by the inflaton's decay.
Confronting our scenario with the current observational data on
the inflationary observables, the light neutrino masses, the
baryon asymmetry of the universe and the gravitino limit on the
reheat temperature, we constrain the strength of the gravitational
coupling to rather large values $(\sim 45-2950)$ and the Dirac
neutrino masses to values between about $1$ and $10~\GeV$. \\ \\
{\scriptsize {\sf PACs numbers: 98.80.Cq, 11.30.Qc, 11.30.Er,
11.30.Pb, 12.60.Jv} \hfill {\sl\bfseries Published in} {\sl Phys.
Rev. D} {\bf 86}, 023523 (2012)}

\end{abstract}\pagestyle{fancyplain}

\maketitle

\rhead[\fancyplain{}{ \bf \thepage}]{\fancyplain{}{\sl Non-Minimal Chaotic Inflation,
Peccei-Quinn Phase Transition and non-Thermal Leptogenesis}}
\lhead[\fancyplain{}{\sl \leftmark}]{\fancyplain{}{\bf
\thepage}} \cfoot{}

\section{Introduction}

There is recently a wave of interest in implementing
\emph{non-minimal chaotic inflation} (nMCI) within both a
non-\emph{supersymmetric} (SUSY) \cite{sm1,nmchaotic,wmap3,
love,nmi1,nmi,ld,unitarizing} and a SUSY
\cite{linde1,linde2,nmN,nmH, nmSUGRA} framework. The main idea is
to introduce a large non-minimal coupling of the inflaton field to
the curvature scalar, $\rcc$. After that, one can make a
transformation -- from the \emph{Jordan frame} (JF) to the
\emph{Einstein} (EF) one -- which flattens the potential
sufficiently to support nMCI. The implementation of this mechanism
within \emph{supergravity} (SUGRA) has been greatly facilitated
after the developed \cite{linde2} supercorformal approach to
SUGRA. In particular, it is shown that the frame function can be
related to a logarithmic type K\"ahler potential which ensures
canonical kinetic terms for the scalars of the theory and
incorporates an holomorphic function, $F$, which expresses the
non-minimal coupling of the inflaton field to $\rcc$. Until now,
the proposed models \cite{linde2, nmN, nmH, nmSUGRA} of nMCI
within SUGRA are constructed coupling quadratically the inflaton
superfield with another one in the superpotential -- leading
thereby to a quartic potential -- and adopting a quadratic term
for it in $F$.

In this paper we propose a novel realization of nMCI within SUGRA,
according to which the inflaton superfield is coupled linearly to
another superfield in the superpotential of the model. As a
consequence, a quadratic potential for the inflaton arises which
supports nMCI, if the inflaton develops a linear coupling to
$\rcc$. Actually, this set-up represents the SUSY implementation
of the model of \FHI with $n=-1$ introduced in \cref{nmi}. In
contrast to earlier models \cite{chaotic1, linde2} which relied on
the same superpotential term -- see also \cref{Takahashi} --, no
extra shift symmetry is imposed on the K\"alher potential. The
resulting mass of the inflaton lies at the intermediate scale and
the inflationary observables are principally similar to those of
\FHI with quartic -- not quadratic -- potential and therefore, in
excellent agreement with the current observational data
\cite{wmap}.

The inflationary model can be nicely embedded in a modest
phenomenological extension of the \emph{minimal supersymmetric
standard model} (MSSM) which incorporates a resolution of the
strong CP problem \cite{pq} via a \emph{Peccei-Quinn} (PQ)
symmetry. Note that there is an increasing interest \cite{Covi,
Baerax} in such models at present, since they provide us with two
additional \emph{cold dark matter} (CDM) candidates (axino and
axion) beyond the lightest neutralino. In our model, a \emph{PQ
phase transition} (PQPT), tied on renormalizable \cite{goto, nmN}
superpontential terms, can follow nMCI generating in addition, the
$\mu$ term of MSSM and intermediate masses for the \emph{right
handed} (RH) [s]neutrinos, $\sni$ [$\ssni$]. As a consequence, the
light neutrino masses can be explained through the well-known
see-saw mechanism \cite{seesaw} provided that no large hierarchies
occur in the Dirac neutrino masses. The possible formation
\cite{sikivie} of disastrous domain walls can be avoided
\cite{georgi, cdmon} by introducing extra matter superfields
without jeopardizing the gauge unification of MSSM. The appearance
of a Lagrangian quatric coupling of the inflaton ensures its decay
to $\ssni$, whose the subsequent out-of-equilibrium decays can
generate the \emph{Baryon Asymmetry of the Universe} (BAU) via
\emph{non-thermal leptogenesis} (nTL) \cite{inlept}, consistent
with the present data on neutrino data \cite{Expneutrino, Lisi}.
Our model favors mostly quasi-degenerate $\sni$ -- as in
\cref{Asaka} -- which enhances the contribution from the
self-energy corrections to leptonic asymmetries, without
jeopardizing the validity of the relevant perturbative results,
though. The constraints arising from BAU and the gravitino ($\Gr$)
limit \cite{gravitino, brand, kohri} on the reheat temperature
can be met provided that the masses of $\Gr$ lie in the
multi-$\TeV$ region.

In Sec.~\ref{fhim} we present the basic ingredients of our model,
Sec.~\ref{fhi} describes the inflationary scenario, and we outline
the mechanism of nTL in Sec.~\ref{pfhi}. We then restrict the
model parameters in Sec.~\ref{cont} and summarize our conclusions
in Sec.~\ref{con}. Throughout the text, the subscript of type
$,\chi$ denotes derivation \emph{with respect to} (w.r.t) the
field $\chi$ (e.g., $_{,\chi\chi}=\partial^2/\partial\chi^2$);
charge conjugation is denoted by a star and brackets are, also,
used by applying disjunctive correspondence.

\section{Model Description}\label{fhim}

We focus on a PQ invariant extension of MSSM, which is augmented
with {\small \sf  (i)} two superfields ($P$ and $\bP$) which are
necessary for the implementation of nMCI; {\small \sf (ii)} three
superfields ($S, \Xb$ and $\bX$) involved in the spontaneous
breaking of the PQ symmetry, $U(1)_{\rm PQ}$ {\small \sf  (iii)}
three RH neutrinos, $\sni$, which are necessitated for the
realization of the see-saw mechanism; {\small \sf  (iv)} {\rm n}
-- to be determined below -- pairs  of $SU(3)_{\rm C}$  triplets
and antitriplets superfields, $\bar D_{\rm a}$ and $D_{\rm a}$
respectively, (${\rm a} = 1, ..., {\rm n}$) in order to avoid the
formation of domain walls -- c.f. Ref.~\cite{georgi,cdmon} -- and
{\small \sf  (v)} an equal number of pairs of $SU(2)_{\rm L}$
doublet superfields, $\bar h_{\rm a}$ and $h_{\rm a}$ in order to
restore gauge coupling unification at one loop -- see below.
Besides the superfields in the points (iv) and (v), all the others
are singlets under the \emph{Standard Model} (SM) gauge group
${G_{\rm SM}}= SU(3)_{\rm C}\times SU(2)_{\rm L}\times U(1)_{Y}$.
Besides the (color) anomalous $U(1)_{\rm PQ}$, the model also
possesses an anomalous $R$ symmetry $U(1)_{R}$ the baryon number
symmetry $U(1)_B$ and two accidental symmetries $U(1)_{D}$ and
$U(1)_{h}$. The representations under ${G_{\rm SM}}$, and the
charges under the global symmetries of the various matter and
Higgs superfields are listed in Table~\ref{tab1}. Note that the
lepton number is not conserved in our model.

In particular, the superpotential, $W$, of our model can be split
into four parts:
\beq W=W_{\rm MSSM}+W_{\rm DW}+W_{\rm CPQ}+W_{\rm NR},
\label{Wtot}\eeq
which are analyzed in the following:

\paragraph{} $W_{\rm MSSM}$ is the part of $W$ which contains the
usual terms -- except for the $\mu$ term -- of MSSM, supplemented
by Yukawa interactions among the left-handed leptons and $\sni$:
\bea  \nonumber W_{\rm MSSM}& =& h_{Dij} {d}^c_i {Q}_j \hd + h_{Uij}
{u}^c_i {Q}_j \hu\\ &+&h_{Eij} {e}^c_i {L}_j \hd+ h_{Nij} \sni L_j
\hu. \label{wmssm}\eea
Here, the group indices have been suppressed and summation over
the generation indices $i$ and $j$ is assumed; the $i$-th
generation $SU(2)_{\rm L}$ doublet left-handed quark and lepton
superfields are denoted by $Q_i$ and $L_i$ respectively, and
the $SU(2)_{\rm L}$ singlet antiquark [antilepton] superfields by
$u^c_i$ and ${d_i}^c$ [$e^c_i$ and $\sni$] respectively. The
electroweak $SU(2)_{\rm L}$ doublet Higgs superfield, which
couples to the up [down] quark superfields, is denoted by $\hu$
[$\hd$].

\paragraph{} $W_{\rm DW}$ is the part of $W$ which gives intermediate scale
masses via $\vev{\bX}$ -- see below -- to $\bar D_{\rm a}-D_{\rm
a}$ and $\bar h_{\rm a}-h_{\rm a}$. Namely,
\beq\label{Wdw} W_{\rm DW} =\lambda_{D\rm a}\bX\bar D_{\rm a}
D_{\rm a}+\lambda_{h\rm a}\bX \bar h_{\rm a} h_{\rm a}. \eeq
Here, we chose a basis in the $\bar D_{\rm a}-D_{\rm a}$ and $\bar
h_{\rm a}-h_{\rm a}$ space where the coupling constant matrices
$\lambda_{D\rm a}$ and $\lambda_{h\rm a}$ are diagonal. Although
these matter fields acquire intermediate scale masses after the PQ
breaking, the unification of the MSSM gauge coupling constants is
not disrupted at one loop. In fact, if we estimate the
contribution of $\bar D_{\rm a}, D_{\rm a},$ and $\bar h_{\rm a}$
and $h_{\rm a}$ to the coefficients $b_1,~b_2,$ and $b_3$,
controlling \cite{Jones} the one loop evolution of the three gauge
coupling constants $g_1, g_2,$ and $g_3$, we find that the
quantities $b_2-b_1$ and $b_3-b_2$ (which are \cite{Jones} crucial
for the unification of $g_1, g_2,$ and $g_3$) remain unaltered.

\paragraph{} $W_{\rm CPQ}$ is the part of $W$ which is
relevant for nMCI, the spontaneous breaking of ${U(1)}_{\rm PQ}$,
the decay of the inflaton and the generation of the masses of
$\sni$'s and the $\mu$ term of MSSM. It takes the form
\beq W_{\rm CPQ}= m \bP P+\la S\lf \Xb\bX-\Mpq^2
\rg+\ld_{i\nu^c}\Xb \nu_i^{c2}, \label{Whi}\eeq
where $\Mpq=f_a/2$ with $f_a\simeq\lf10^{10}-10^{12}\rg~{\rm GeV}$
being the axion decay constant which coincides with the PQ
breaking scale. The parameters $\la$ and $f_a$ can be made
positive by field redefinitions. From the terms in the \emph{right
hand side} (RHS) of \Eref{Whi} we note that the imposed symmetries
disallow renormalizable terms mixing $\bP$ with some other
superfields, which avoids undesirable instabilities faced in
\cref{nmN}.

\paragraph{} $W_{\rm NR}$ is the part of $W$ which
contains its non-renormalizable terms. Namely, we have
\beq W_{\rm NR}=\ld_i{\bP S \Phi
\sni\over\mP}+\ld_P{P\bP\bX\Xb\over\mP}+\lm{\bX^2\hu\hd\over\mP},\label{Wnr}\eeq
where $m_{\rm P}\simeq 2.44\cdot 10^{18}~{\rm GeV}$ is the reduced
Planck scale. The first term in the RHS of \Eref{Wnr} helps
accomplish sufficiently low reheat temperature and leads to the
production of $\ssni$'s as dictated by nTL  -- see \Sref{lept}.
Finally, the third term provides the $\mu$ term of MSSM -- see
below.

\begin{table}[!t]
\caption{\normalfont Superfield Content of the Model}
\begin{tabular}{c@{\hspace{0.4cm}}c@{\hspace{0.4cm}}c@{\hspace{0.4cm}}c@
{\hspace{0.4cm}}c@{\hspace{0.4cm}}c@{\hspace{0.4cm}}c} \toprule
{Super-}&{Representations}&\multicolumn{5}{c}{Global
Symmetries}\\
{fields}&{under $G_{\rm SM}$}&$R$&PQ &{$B$}&{$D$}&{$h$} \\\colrule
\multicolumn{7}{c}{Matter Fields}\\\colrule
{$L_i$} &{$({\bf 1, 2}, -1/2)$}& $0$ & $-1$ &$0$ &$0$ &$0$
\\
{$e^c_i$} & {$({\bf 1, 1}, 1)$} &$2$&{$-1$}&{$0$}&$0$ &$0$
\\
{$\sni$} & {$({\bf 1, 1}, 0)$} &$2$&{$-1$}&{$0$}&$0$ &$0$
\\
{$Q_i$} &{$({\bf 3, 2}, 1/6)$}& $1$ & $-1$ &$1/3$&$0$ &$0$
 \\
{$u^c_i$} & {$({\bf \bar 3, 1}, -2/3)$} &$1$ &{$-1$}&$-1/3$&$0$
&$0$
\\
{$d^c_i$} & {$({\bf \bar 3, 1}, 1/3)$} &$1$ &{$-1$}&$-1/3$&$0$
&$0$
\\ \colrule
\multicolumn{7}{c}{Extra Matter Fields} \\ \colrule
$D_{\rm a}$&$({\bf 3, 1}, -1/3)$& $1$ & $1$ &$0$&$1$ &$0$\\
$\bar D_{\rm a}$&{$({\bf \bar 3, 1}, 1/3)$}& $1$
&$1$&$0$&$-1$ &$0$\\
$h_{\rm a}$&$({\bf 1, 2}, 1/2)$ & $1$ & $1$ &$0$&$0$ &$1$\\
$\bar h_{\rm a}$&$({\bf 1, 2}, -1/2)$ & $1$ & $1$ &$0$&$0$ &$-1$
\\ \colrule
\multicolumn{7}{c}{Higgs Fields} \\\colrule
{$\hd$} & {$({\bf 1, 2}, -1/2)$}&$2$ &$2$ &$0$&$0$ &$0$ \\
{$\hu$} & {$({\bf 1, 2}, 1/2)$}&$2$ &$2$ &$0$&$0$ &$0$ \\ \hline
{$S$} &{$({\bf 1, 1}, 0)$}&{$4$}&{$0$} &{$0$}&$0$ &$0$\\
{$\Xb$}&$({\bf 1, 1}, 0)$&{$0$}&{$2$}&{$0$}&$0$ &$0$\\
{$\bX$}&$({\bf 1, 1}, 0)$&{$0$}&{$-2$}&{$0$}&$0$ &$0$\\\colrule
{$P$} &{$({\bf 1, 1}, 0)$}&{$6$}&{$1$} & {$0$}&$0$ &$0$\\
{$\bar P$} &{$({\bf 1, 1}, 0)$}&{$-2$}&{$-1$} & {$0$}&$0$ &$0$
\\ \botrule
\end{tabular}
\label{tab1}
\end{table}

To get an impression for the role that each term in the RHS of
\eqss{Wdw}{Whi}{Wnr} play, we display the SUSY potential, $V_{\rm
SUSY}$, induced from the following part of $W$
\beq W_{\rm CI}=W_{\rm CPQ}+W_{\rm DW},\label{Wci}\eeq which turns
out to be
\beqs\bea \nonumber V_{\rm SUSY}&= &m^2\lf|P|^2+|\bP|^2\rg+
\left|\ld_{i\nu^c}\tilde\nu_i^{c2}+\la S\bX\right|^2\\&+&
\nonumber 4\ld_{i\nu^c}^2|\sn_i\Xb|^2+\la^2\left|\bX
\Xb-\Mpq^2\right|^2 \\\nonumber &+& \left|\la S\Xb+\ld_{D{\rm
a}}\dba\da+\ld_{h{\rm a}}\hba\ha\right|^2\\&+& \nonumber
\ld_{D{\rm a}}^2\lf|\dba|^2+|\da|^2\rg|\bX|^2\\&+& \ld_{h{\rm
a}}^2\lf|\hba|^2+|\ha|^2\rg|\bX|^2, \label{VF}\eea
where the complex scalar components of the superfields $P, \bP, S,
\bX, \Xb, \dba, \da, \hba,$ and $\ha$  are denoted by the same
symbol as the corresponding superfields. From \Eref{Wci} and
assuming \cite{goto} canonical \Ka\ for the hidden sector fields,
we can also derive the soft SUSY-breaking part of the inflationary
potential which reads:
\bea \nonumber V_{\sft}&=&m^2_{\phi^\al}\phi^\al\phi^*_\al
+\Big(mBP\bar P-{\rm a_T}\la S\Mpq^2\\
&+&\ld_{D\rm a} A_{D\rm a}\bX \bar D_{\rm a} D_{\rm a}+\ld_{h\rm
a} A_{h\rm a}\bX \bar h_{\rm a} h_{\rm a}\nonumber\\ &+&\la
A_aS\Xb\bX+\ld_{i\nu^c}A_{i\nu^c}\Xb\nu^c_i +{\rm h.c.}\Big)
\label{Vsoft}\eea\eeqs
where $m_{\phi^\al}$, with \beq \phi^\al=P, \bP, S, \bX, \Xb,
\ssni, \dbma, \dma, \hbla, \hla\label{fas}\eeq $A_a$,
$A_{i\nu^c}$, $A_{D\rm a}$, $A_{h\rm a}$, $B$ and ${\rm a_T}$ are
soft SUSY-breaking mass parameters of order $1~\TeV$. From the
potential in \eqs{VF}{Vsoft}, we find that the SUSY vacuum lies at
\beqs\bea&& \vev{P}=\vev{\bP}=\vev{\ssni}=0,\label{vevs1}
\\ && \vev{\dma}=\vev{\dbma}=\vev{\hla}=\vev{\hbla}=0,\>\>\>
\label{vevs3}\eea and \bea\vev{S}={|A_a|+|{\rm
a_T}|\over2\la},~|\langle\phi_{\Xb}\rangle|=2|\vev{\Xb}|=2|\vev{\bX}|=f_a,\label{vevs2}
\eea\eeqs
where the resulting $\vev{S}$ is of the order of $\TeV$ -- cf.
\cref{goto} -- and we have introduced the canonically normalized
scalar field $\phi_{\Xb}=2\Xb=2\bX$. Also, we use the subscripts
$k=1,2,3$ and $l=1,2$ to denote the components of $\da,\dba$ and
$\ha,\hba$, respectively. Note that, since the sum of the
arguments of $\vev{\bX}$, $\vev{\Xb}$ must be $0$, $\bX$ and $\Xb$
can be brought to the real axis by an appropriate PQ
transformation. After the spontaneous breaking of $U(1)_{\rm PQ}$,
the third term in \Eref{Whi} generates intermediate scale masses,
$\mrh[i]$ for the $\sni$'s and, thus, seesaw masses \cite{seesaw}
for the light neutrinos -- see \Sref{pfhi}. The third term in the
RHS of \Eref{Wnr} leads to the $\mu$ term of MSSM, with
$|\mu|\sim\lambda_\mu\left|\vev{\bX}\right|^2/m_{\rm P}$, which is
of the right magnitude if $\left|\vev{\bX}\right|=f_a/2\simeq
5\cdot 10^{11}~{\rm GeV}$, $\lambda_\mu\sim(0.001-0.01)$. Finally,
since $\vev{\bX\Xb}/\mP=\Mpq^2/\mP\ll m\simeq10^{16}~\GeV$ -- see
\Sref{fhi2} and \ref{num1} -- the second term in the RHS of
\Eref{Wnr} has no impact on our results.

Nonetheless, $W_{\rm CI}$ also gives rise to a stage of \FHI
within SUGRA, if it is combined with a suitable  K\"ahler
potential, $\Khi$, related to the frame function, $\Ohi$ via
\beq\Khi=-3\mP^2\ln\lf-{\Ohi/3}\rg.\label{Om}\eeq
In JF a specific form of $\Ohi$'s -- see \cref{linde2, nmN} --
ensures canonical kinetic terms of the fields involved and a
non-minimal coupling of the inflaton to $\rcc$ represented by an
holomorphic function $F(P)$. Going from JF to EF, and expanding
the EF potential, $\what V$, along a stable direction -- usually
with all the fields besides inflaton placed at the origin --
$\what V$ takes the simple form
\beq\what V_{\rm CI0}\simeq V_{\rm
SUSY}/f\lf\sg\rg^2,\label{Vhi1}\eeq
where $\sg=\sqrt{2}|P|$ and $f$ can be found expanding $\Ohi$.
Vanishing of the non-inflaton fields ensures, also, the
elimination of some extra kinetic terms for scalars from the
auxiliary vector fields -- see \cref{linde1,linde2,nmN}.

Let us emphasize here that the coupling of $P$ to $\bar P$ is
crucial in order to obtain the simple form of $\what V_{\rm CI0}$
in \Eref{Vhi1}, since only terms including derivatives of $W_{\rm
CI}$ w.r.t $\bar P$ survive in the EF SUGRA potential -- see
\Sref{fhi1}. This fact ensures the appearance of just one dominant
power of $\sigma$ in the numerator of the SUGRA scalar potential.
Such a construction is not possible, e.g., for a superpotential
term of the form $m P^2$. Applying the strategy, described above
\Eref{Vhi1} in our case, we can observe that along the direction
\beqs\bea && \nonumber\th=\bP=S=\bX=\Xb=\ssni=\\
&& \dbma=\dma=\hbla=\hla=0,\label{inftr}\eea
with $\th={\sf arg}P$, $V_{\rm SUSY}$ in \Eref{VF} becomes
\beq V_{\rm SUSY}={1\over2}m^2\sg^{2}+\la^2 \Mpq^4.\eeq\eeqs
Clearly, for $\sg\gg f_a$, $V_{\rm SUSY}$ tends to a quadratic
potential which can be flattened, according to \Eref{Vhi1}, if $f$
is mainly proportional to $\sg$, i.e., if $F$ is a linear function
of $P$ with a sizable coupling constant $\ck$. Therefore, we are
led to adopt the following frame function:
\beqs\beq \Ohi=-3+{\phi^\al\phi^*_\al \over\mP^2}-
{\kx\over\mP^4}|\bP|^4-\big({F}\lf P\rg+{F}^*\lf P^*\rg\big),
\label{minK}\eeq
with $\phi^\al$'s defined in \Eref{fas} and the non-minimal
gravitational coupling \beq {F}=3\ck P/\sqrt{2}\mP, \label{Fdef}
\eeq\eeqs which breaks explicitly the imposed $R$ and PQ
symmetries during nMCI. In \Eref{minK} the coefficients $\kx$ and
$\ck$, for simplicity, are taken real. We remark that we add the
third term in the RHS of \Eref{minK} to cure the tachyonic mass
problem encountered in similar models \cite{linde1, linde2} -- see
\Sref{fhi1}.

%

For $P\ll\mP$, we can show -- see \Sref{lept} -- that an
instability occurs in the PQ system which can drive a PQPT which
leads to the v.e.vs in \Eref{vevs2}. Also, at the SUSY vacuum the
explicit breaking of $U(1)_R\times U(1)_{\rm PQ}$ through
\Eref{Fdef} switches off -- see \Eref{vevs1}. A closer look,
however, reveals that instanton and soft SUSY breaking effects
explicitly break $U(1)_R\times U(1)_{\rm PQ}$ to
$\mathbb{Z}_2\times \mathbb{Z}_{2({\rm n}-6)}$, as can be deduced
from the solutions of the system
\beq \label{expb} 4r=0~\lf\mbox{\ftn\sf
mod}~2\pi\rg\>\>\>\mbox{and}\>\>\> 2({\rm
n}-6)p-12r=0~\lf\mbox{\ftn\sf mod}~2\pi\rg, \eeq
where $r$ and $p$ are the phases of a $U(1)_{R}$ and $U(1)_{\rm
PQ}$ rotation respectively. Here, we take into account that the
$R$ charge of $W$ and, thus, of all the soft SUSY breaking term is
4 and that the sum of the $R$ [PQ] charges of the $SU(3)_{\rm C}$
triplets and antitriplets is $-12$~[$2({\rm n}-6)$]. Note that no
loop-induced PQ-violating term -- as this appearing in the first
paper of \cref{nmSUGRA} -- is detected in our case. It is then
important to ensure that $\mathbb{Z}_2\times \mathbb{Z}_{2({\rm
n}-6)}$ is not spontaneously broken by $\vev{\Xb}$ and
$\vev{\bX}$, since otherwise cosmologically disastrous domain
walls are produced \cite{sikivie} during PQPT. This goal can be
accomplished by adjusting conveniently the number n of $\dba-\da$
and $\hba-\ha$ -- see \Tref{tab1}. Indeed, when ${\rm n}=5$ or $7$
we obtain $2p=0~\lf\mbox{\ftn\sf mod}~2\pi\rg$ and therefore,
$\mathbb{Z}_2\times \mathbb{Z}_{2({\rm n}-6)}$ is not
spontaneously broken by $\vev{\Xb}$ and $\vev{\bX}$. The residual
unbroken $\mathbb{Z}_2$ subgroup of $U(1)_{\rm PQ}$ can be
identified with the usual matter parity of MSSM -- see \Tref{tab1}
-- which prevents the rapid proton decay and ensures the stability
of the \emph{lightest SUSY particle} (LSP).

\section{The Inflationary Epoch}\label{fhi}

In \Sref{fhi1} we describe the salient features of our
inflationary model and in \Sref{fhi2} we extract the inflationary
observables.

\subsection{\bf\scshape Structure of the Inflationary
Potential}\label{fhi1}

The EF F--term (tree level) SUGRA scalar potential, $\Vhio$, of
our model is obtained from $W_{\rm CI}$ in Eq.~(\ref{Wci}) and
$\Khi$ in \eqs{Om}{minK} by applying \cite{linde1}
\beqs\bea \Vhio=e^{\Khi/\mP^2}\left(K^{\al\bbet}{\rm F}_\al {\rm
F}_\bbet-3\frac{\vert W_{\rm
CI}\vert^2}{\mP^2}\right),\label{Vsugra} \eea with \bea
K_{\al\bbet}={\Khi_{,\phi^\al\phi^{*\bbet}}},\>\>\>
K^{\bbet\al}K_{\al\bar \gamma}=\delta^\bbet_{\bar \gamma}\eea
and \bea {\rm F}_\al=W_{\rm CI,\phi^\al} +K_{,\phi^\al}W_{\rm
CI}/\mP^2, \eea\eeqs
where the $\phi^\al$'s are given in \Eref{fas}. From the resulting
$\Vhio$, we can deduce that along the field directions in
\Eref{inftr},
\beqs\bea \label{Vhi}\Vhio =\frac{m^2\mP^2\xsg^2+ 4 \la^2
\Mpq^4/\mP^4}{2f^2}\simeq{m^2\mP^2\xsg^2\over2f^2},\eea
where $\xsg=\sg/\mP$ and, according to the general recipe
\cite{linde1, linde2, nmN}, the function
\bea f=1+\ck\xsg-{\xsg^2/6} \eea\eeqs
expresses the non-minimal coupling of $\sg$ to $\rcc$ in JF. From
\Eref{Vhi}, we can verify that for $\ck\gg1$ and
$\xsg\ll\sqrt{6}$, $\Vhi$ develops a plateau since $\Mpq\ll\mP$ --
see \Sref{fhi2}. Along the trajectory in \Eref{inftr}, we can
estimate the constant potential energy density
\beqs\beq\label{Vhio} \Vhio=
{m^2\sg^2\over2f^2}\simeq{m^2\mP^2\over2\ck^2},\eeq and the
corresponding Hubble parameter \beq \He_{\rm
CI}={\Vhio^{1/2}\over\sqrt{3}\mP}\simeq{m\over\sqrt{6}\ck}
\cdot\eeq\eeqs

In order to check the stability of the direction in \Eref{inftr}
w.r.t the fluctuations of the various fields, we expand them in
real and imaginary parts according to the prescription
\beqs\beq P={\sg e^{i\th}\over\sqrt{2}}~~~\mbox{and}~~~X=
{\chi_1+i\chi_2\over\sqrt{2}},\label{cannor} \eeq
where \beq X=\bP, S, \bX, \Xb, \ssni, \dbma,\dma,\hbla,\hla\eeq
and \beq \chi=\bar p, s,\bar \phi,  \phi, \nu_i, \dbma, \dma,
\hbla, \hla, \eeq \eeqs respectively. Along the trajectory in
\Eref{inftr} we find
\beqs\bea \lf K_{\al\bbet}\rg=\diag\lf
J^2,\underbrace{1/f,...,1/f}_{7+10{\rm n} ~\mbox{\ftn
elements}}\rg, \eea where \bea \label{VJe}
J=\sqrt{\frac{1}{f}+{3\over2}\mP^2\left({f_{,\sg}\over
f}\right)^2}={\sqrt{2+3\ck^2}\over\sqrt{2} f}\simeq
\sqrt{3\over2}{1\over\xsg}\cdot\>\>\eea\eeqs
Consequently, we can introduce the EF canonically normalized
fields, $\se, \widehat \th$ and $\widehat \chi$, as follows -- cf.
\cref{linde1,linde2,nmN,nmH}:
\beq K_{\al\bbet}\dot\phi^\al
\dot\phi^{*\bbet}={1\over2}\lf\dot{\what\sigma}^{2}+\dot{\what
\th}^{2}\rg+{1\over2}\sum_\chi\lf\dot{\what\chi}_1^2+\dot{\what\chi}_2^2\rg,\eeq
where the dot denotes derivation w.r.t the JF cosmic time, $t$ and
the hatted fields are defined as follows
\beq  {d\widehat \sg\over d\sg}=J,\>\>\> \widehat \th\simeq J\sg
\th \>\>\>\mbox{and}\>\>\>\what \chi \simeq\frac{\chi}{\sqrt{f}}
\cdot\eeq
Note that $\dot{\widehat \th}\simeq J\sg\dot \th$ since
$J\sg\simeq\sqrt{3/2}\mP$ -- see \Eref{VJe}. On the other hand, we
can show that during a stage of slow-roll nMCI, $\dot{\widehat
\chi}\simeq\dot \chi/\sqrt{f}$ since the quantity $\dot
f/2f^{3/2}\chi$, involved in relating $\dot{\widehat \chi}$ to
$\dot \chi$, turns out to be negligibly small compared with
$\dot{\what\chi}$. Indeed, the $\what \chi$'s acquire effective
masses $m_{\what \chi}\gg \Hhi$ -- see below -- and therefore
enter a phase of oscillations about $\what \chi=0$ with reducing
amplitude. Neglecting the oscillating part of the relevant
solutions, we find
\beq \chi\simeq\what
\chi_{0}\sqrt{f}e^{-2\Ne/3}\>\>\mbox{and}\>\>\>
\dot{\what\chi}\simeq-2\chi_{0}\sqrt{f}\Hhi\what\eta_\chi
e^{-2\Ne/3},\label{xdx}\eeq
where $\what \chi_{0}$ represents the initial amplitude of the
oscillations, $\what\eta_\chi=m^2_{\what\chi}/3\Hhi$ and we assume
$\dot{\what \chi}(t=0)=0$. Taking into account the approximate
expressions for $\dot\sg$ and the slow-roll parameter
$\widehat\epsilon$, which are displayed in \Sref{fhi2}, we find
\beq -{\dot f\over2f^{3/2}}\chi={\ck\what\epsilon\He_{\rm
CI}^2\over m_{\what \chi}^2}\ \dot{\what\chi}\ll
\dot{\what\chi}.\label{ff32}\eeq

\renewcommand{\arraystretch}{1.4}

\begin{table}[!t]
\caption{\normalfont The mass spectrum of the model during nMCI}
\begin{tabular}{c|@{\hspace{0.1cm}}c@{\hspace{0.1cm}}|@{\hspace{0.1cm}} c}\toprule
{Fields} &{Eingestates} & {Mass Squared}\\ \colrule
\multicolumn{3}{c}{Bosons}\\ \colrule
$1$ real scalar &$\what \th$ & $\ck m^2\xsg/f^3J^2\simeq4H_{\rm CI}^2$\\
$2$ real scalars &$\what{\bar p}_1,\what{\bar p}_2$ & $m^2\lf2\kx\ck\xsg^3+(-\ck\xsg+\right.$\\
&& $\left.(6\kx-1)\ck^2\xsg^2)/2J^2f^2\rg/f^2$\\
$2$ real scalars &$\what{s}_1,\what{s}_2$ &
$\lf 2\la\Mpq^4/\mP^2+m^2\xsg^2\rg/3f^2$\\
$6$ real scalars &$\what{\nu}_{1i},\what{\nu}_{2i}$ &
$\lf 2\la\Mpq^4/\mP^2+m^2\xsg^2\rg/f^2$\\
$4$ real &
${\what{\bar\phi}_1\pm\what{\phi}_1\over\sqrt{2}},$ & $\lf m^2\xsg^2/3\pm\la^2\Mpq^2f\rg/f^2$\\
scalars & ${\what{\bar\phi}_2\pm\what{\phi}_2\over\sqrt{2}}$ &\\
$6{\rm n}$ real scalars &$\what{D}_{1k{\rm a}},\what{D}_{2k{\rm
a}}$ &
$\lf 2\la\Mpq^4/\mP^2+m^2\xsg^2\rg/f^2$\\
$6{\rm n}$ real scalars &$\what{\bar D}_{1k{\rm a}},\what{\bar
D}_{2k{\rm a}}$ &
$\lf 2\la\Mpq^4/\mP^2+m^2\xsg^2\rg/f^2$\\
$4{\rm n}$ real scalars &$\what{h}_{1l{\rm a}},\what{h}_{2l{\rm
a}}$ &
$\lf 2\la\Mpq^4/\mP^2+m^2\xsg^2\rg/f^2$\\
$4{\rm n}$ real scalars &$\what{\bar h}_{1l{\rm a}},\what{\bar
h}_{2l{\rm a}}$ & $\lf
2\la\Mpq^4/\mP^2+m^2\xsg^2\rg/f^2$\\\colrule
\multicolumn{3}{c}{Fermions}\\ \colrule
$2$ Weyl spinors & ${\what{\psi}_{\bar P}\pm \what{\psi}_{P}\over\sqrt{2}}$& $m^2(6+\xsg^2)^2/36f^2$\\
\botrule
\end{tabular}
\label{tab2}
\end{table}

The masses that the various scalars acquire during nMCI are
presented in \Tref{tab2}. To this end, we expand $\Vhio$ in
\Eref{Vsugra} to quadratic order in the fluctuations around the
direction of \Eref{inftr}. As we observe from the relevant
eigenvalues of the mass-squared matrices, no instability -- as the
one found in \cref{nmN} -- arises in the spectrum. In particular,
it is evident that $\kx\gtrsim1$ assists us to achieve positivity
of the mass-squared associated with the scalars $\what{\bar
p}_{1,2}$, $m_{\widehat{\bar p}}^2$ -- in accordance with the
results of \cref{linde1, linde2}. It is remarkable that
mass-squared corresponding to $\ssni, \dma, \dbma, \hla, \hbla$
are independent of the relevant superpotential couplings
$\ld_{i\nu^c},\ld_{D{\rm a}}$ and $\ld_{h{\rm a}}$. We have also
numerically verified that the various masses remain greater than
$\Hhi$ during the last $50-60$ e-foldings of nMCI, and so any
inflationary perturbations of the fields other than the inflaton
are safely eliminated.

In \Tref{tab2} we also present the masses squared of
chiral fermions of the model along the direction of \Eref{inftr}.
Inserting these masses into the well-known Coleman-Weinberg
formula \cite{cw}, we can find the one-loop radiative corrections,
$V_{\rm rc}$, in our model which can be written as
\bea \nonumber V_{\rm rc}&=& {1\over64\pi^2}\lf m_{\widehat
\th}^4\ln{m_{\widehat p}^2\over\Lambda^2}+2m_{\widehat{\bar
p}}^4\ln{m_{\widehat{\bar p}}^2\over\Lambda^2}+
2m_{\widehat{s}}^4\ln{m_{\widehat{s}}^2\over\Lambda^2}\right.\\
&+&6m_{\widehat{\nu}}^4\ln{m_{\widehat{\nu}}^2\over\Lambda^2}
+2m_{\widehat{\phi}_+}^4\ln{m_{\widehat{\phi}_+}^2\over\Lambda^2}+
2m_{\widehat{\phi}_-}^4\ln{m_{\widehat{\phi}_-}^2\over\Lambda^2} \nonumber\\
&+&\left.20{\rm n}\
m_{\widehat{D}}^4\ln{m_{\widehat{D}}^2\over\Lambda^2}
-4m_{\widehat{\psi}_\pm}^4\ln{m_{\widehat{\psi}_\pm}^2\over\Lambda^2}\right),
\label{Vrc}\eea
where $\Ld=\mP/\ck$ is the cutoff scale of the effective theory --
see \Sref{cont1} -- and the involved above masses squared
$m^2_{\widehat \th},~m^2_{\widehat{\bar p}},~m^2_{\widehat{s}},$
$m^2_{\widehat{\nu}},~m^2_{\widehat{\phi}_\pm},m^2_{\widehat{D}}$
and $m^2_{\widehat{\psi}_\pm}$ are equal to the ones listed in the
third column of \Tref{tab2} from top to the bottom. Note that the
masses squared of all the extra matter fields are equal to
$m^2_{\widehat{D}}$. Based on the one-loop corrected EF potential
\beq \Vhi=\Vhio+V_{\rm rc},\label{Vhic} \eeq
we can proceed to the analysis of \FHI in EF, employing the
standard slow-roll approximation \cite{review}. It can be shown
\cite{general} that the results calculated this way are the same
as if we had calculated them using the non-minimally coupled
scalar field in JF. As expected and verified numerically, $V_{\rm
rc}$ does not affect the inflationary dynamics and predictions, in
the major part of the allowed parameter space -- see \Sref{num1}
-- since the inflationary path already possesses a slope at the
classical level -- see below.


\subsection{\bf\scshape The Inflationary Observables}\label{fhi2}

According to our analysis above, the universe undergoes a period
of slow-roll nMCI, which is determined by the condition -- see
e.g. \cref{review}:
\bea{\ftn\sf
max}\{\widehat\epsilon(\sigma),|\widehat\eta(\sigma)|\}\leq1,\nonumber\eea
where\beqs\beq\label{sr1}\widehat\epsilon=
{\mP^2\over2}\left(\frac{\Ve_{{\rm CI},\se}}{\Ve_{\rm
CI}}\right)^2={\mP^2\over2J^2}\left(\frac{\Ve_{{\rm CI},\sigma}}
{\Ve_{\rm CI}}\right)^2\simeq {4\mP^2\over3\ck^2\sg^2},\eeq
and \bea \widehat\eta &=& m^2_{\rm P}~\frac{\Ve_{{\rm
CI},\se\se}}{\Ve_{\rm CI}}\nonumber  ={\mP^2\over J^2}\left(
\frac{\Ve_{{\rm CI},\sigma\sigma}}{\Ve_{\rm CI}}-\frac{\Ve_{{\rm
CI},\sigma}}{\Ve_{\rm CI}}{J_{,\sg}\over
J}\right)\\&\simeq&-{4\mP/3\ck\sg}. \label{sr2}\eea\eeqs
Here, we employ \eqs{Vhio}{VJe} and the following approximate
relations:
\beq \widehat V_{\rm CI,\sg}\simeq
{m^2\mP\over\ck^3\xsg^2}\>\>\>\mbox{and}\>\>\>\widehat V_{\rm
CI,\sg\sg}\simeq-{2m^2\over\ck^3\xsg^3}\cdot\eeq
The numerical computation reveals that \FHI terminates due to the
violation of the $\widehat\epsilon$ criterion at $\sg=\sgf$, which
is calculated to be
\beq \widehat\epsilon\lf\sgf\rg=1\>\>\>\Rightarrow\>\>\>
\sgf={2\over \sqrt{3}}{\mP\over\ck}\cdot \label{sgap}\eeq
We note, in passing, that for $\sg\geq\sgf$ the evolution of
$\widehat\sg$ -- or $\sg$ via \Eref{VJe} -- is governed by the
equation of motion
\beq 3\Hhi {d{\what{\sg}}\over d\what t}=-\widehat V_{\rm CI,\hat{\sg}}
\>\Rightarrow\>\dot\sg=-{2\sqrt{2}m\over
3\sqrt{3}}\sqrt{\mP^{3}\over \ck^{3}\sg}, \label{eqf}\eeq
where $\what t$ is the EF cosmic time with $d\what t=\sqrt{f} dt$.
Using \eqs{sr1}{eqf}, we can derive \Eref{ff32}.

The number of e-foldings, $\widehat N_{*}$, that the scale
$k_{*}=0.002/{\rm Mpc}$ suffers during nMCI can be calculated
through the relation
\begin{equation}
\label{Nhi}  \widehat{N}_{*}=\:\frac{1}{m^2_{\rm P}}\;
\int_{\se_{\rm f}}^{\se_{*}}\, d\se\: \frac{\Ve_{\rm CI}}{\Ve_{\rm
CI,\se}}= {1\over\mP^2}\int_{\sigma_{\rm f}}^{\sigma_{*}}\,
d\sigma\: J^2\frac{\Ve_{\rm CI}}{\Ve_{\rm CI,\sigma}},
\end{equation}
where $\sigma_*~[\se_*]$ is the value of $\sg~[\se]$ when $k_*$
crosses the inflationary horizon. Given that $\sgf\ll\sg_*$, we
can write $\sg_*$ as a function of $\widehat{N}_{*}$ as follows
\beq \label{s*}
\widehat{N}_{*}\simeq{3\ck\over4\mP}\lf\sg_*-\sgf\rg\>\>\Rightarrow\>\>\sg_*\simeq{4\widehat
N_*\over 3\ck}\mP\cdot\eeq

The power spectrum $\Delta^2_{\cal R}$ of the curvature
perturbations generated by $\sigma$ at the pivot scale $k_*$ is
estimated as follows
\beq  \label{Prob} \Delta_{\cal R}=\:
{1\over2\pi\mP^2}\,\sqrt{\frac{\Vhi(\sg_*)}{6\, \widehat\epsilon\,
(\sg_*)}}\simeq {m\widehat{N}_{*}\over6\pi\mP\ck}, \eeq
where \Eref{s*} is employed to derive the last equality of the
relation above. Since the scalars listed in \Tref{tab2} are
massive  enough during nMCI, $\Delta_{\cal R}$ can be identified
with its central observational value -- see \Sref{cont} -- with
almost constant $\Ne_*$. The resulting relation reveals that $m$
is to be proportional to $\ck$. Indeed we find
\beq m={6\pi\mP\ck \Delta_{\cal R}/\Ne_*}\>\>\Rightarrow\>\>
m=4.1\cdot10^{13}\ck~\GeV,\label{lan}\eeq
for $\Ne_*\simeq55$. At the same pivot scale, we can also calculate
the (scalar) spectral index, $n_{\rm s}$, its running, $a_{\rm s}$,
and the scalar-to-tensor ratio, $r$, via the relations:
\beqs\baq \label{ns} && n_{\rm s}=\: 1-6\widehat\epsilon_*\ +\
2\widehat\eta_*\simeq1-{2/\widehat N_*}, \>\>\> \\
&& \label{as} \alpha_{\rm s}
=\:{2\over3}\left(4\widehat\eta_*^2-(n_{\rm
s}-1)^2\right)-2\widehat\xi_*\simeq{-2/\widehat N^2_*},\\ &&
\label{rt} r=16\widehat\epsilon_*\simeq{12/\widehat N^2_*},
\eaq\eeqs
where $\widehat\xi=\mP^4 {\Ve_{\rm CI,\widehat\sigma} \Ve_{\rm
CI,\widehat\sigma\widehat\sigma\widehat\sigma}/\Ve_{\rm CI}^2}=
\mP\,\sqrt{2\widehat\epsilon}\,\widehat\eta_{,\sigma}/
J+2\widehat\eta\widehat\epsilon$ and the variables with subscript
$*$ are evaluated at $\sigma=\sigma_{*}$. Comparing the results of
this section with the observationally favored values \cite{wmap},
we constrain the parameters of our model in \Sref{num1}.

\section{Non-Thermal Leptogenesis}\label{pfhi}

A complete SUSY inflationary scenario should specify the
transition to the radiation dominated era and also explain the
origin of the observed BAU consistently with the $\Gr$ constraint.
These goals can be accomplished within our set-up, as we describe
in this section. Namely, the basic features of the
post-inflationary evolution are exhibited in \Sref{lept} and the
topic of nTL in conjunction with the present neutrino data is
analyzed in \Sref{lept2}.

\subsection{\bf\scshape  The General Set-up}\label{lept}

When \FHI is over, the inflaton continues to roll down towards the
SUSY vacuum, \eqss{vevs1}{vevs3}{vevs2}. Note that when
$\xsg\lesssim \sqrt{3}\la\Mpq/m$, one scalar originating from the
superfields $\Xb$ and $\bX$ -- see \Tref{tab1} -- acquires a
negative mass squared triggering thereby the PQPT. As the inflaton
continues its rolling, there is a brief stage of tachyonic
preheating \cite{preheating} which does not lead to significant
particle production \cite{garcia}. Soon afterwards, it settles
into a phase of damped oscillations about the minimum of the
$\Vhio$. Since no gauge symmetry is broken during nMCI, no
superheavy bosons are produced and therefore no particle
production via the mechanism of instant preheating \cite{instant}
occurs. Also, since the inflaton cannot decay via renormalizable
interactions to SM particles, effects of narrow parametric
resonance \cite{preheating} are also absent in our regime.

Nonetheless, the standard perturbative approach to the inflaton
decay provides a very efficient decay rate.  Namely, at the SUSY
vacuum the fields involved acquire the v.e.vs shown in
\eqss{vevs1}{vevs3}{vevs2} giving rise to the mass spectrum
presented in \Tref{tab3}. There we can show the mass, $\msn$, of
the (canonically normalized) inflaton $\what P$ and the masses
$\mrh[i]$ of the RH [s]neutrinos, $\nu_i^c$ [$\ssni$], which play
a crucial role in our scenario of nTL. Note that since
$\vev{\Xb}=\vev{\bX}=\Mpq\ll\mP$, $\vev{\Omega}\simeq-3$ and so
$\vev{f}\simeq1$. Therefore, apart from $\what P$, the EF
canonically normalized field are not distinguished from the JF
ones at the SUSY vacuum. On the other hand, $\what P$ can be
expressed as a function of $P$ through the relation
\beq \label{Jo} {\what P\over P}=\vev{J}\>\>\>\mbox{where}\>\>\>
\vev{J}=\sqrt{1+3\ck^2/2}.\eeq
Making use of \Eref{lan} we can infer that $\msn$ is kept
independent of $\ck$ and almost constant at the level of
$10^{13}~\GeV$. Indeed,
\beq \label{mqa} \msn\simeq{m\over
\vev{J}}=\sqrt{2\over3}{m\over\ck}\simeq 2\sqrt{3}\pi \mP
{\Delta_{\cal R}\over\Ne}\simeq10^{13}~\GeV, \eeq
where the WMAP7 value of $\Delta_{\cal R}$ -- see \Sref{cont1} --
is employed in the last step of the relation above. In the
expressions of the various eigenstates listed in \Tref{tab3}, we
adopt the following abbreviations
\beqs\beq\delta\Xb=\Xb-\Mpq,\>\>\delta\bX=\bX-\Mpq,\eeq and
\beq\psi_{\pm}=\lf\psi_{\bX}\pm\psi_{\Xb}\rg/\sqrt{2},\eeq\eeqs
where $\psi_{x}$ with $x=\what P,P,S,\bX,\Xb, \dbma, \dma, \hbla$,
and $\hla$ denote the chiral fermions associated with the
superfields $\what P,P,S,\bX, \Xb, \dbma, \dma, \hla$, and $\hla$
respectively. The eigenstates $\psi_-$ and $\delta\Xb_{-}$, with
\beq\delta\Xb_{\pm}=\lf\delta\bX\pm\delta\Xb\rg/\sqrt{2},\eeq
contain the components of the axion supermultiplet. Namely axion
[saxion] can be identified with the phase [modulus] of the complex
field $\delta\Xb_{-}$, whereas $\psi_-$ can be interpreted as the
axino. Note that the zero masses of saxion and axino can be
replaced with masses of order $1~\TeV$ if we take into account the
soft SUSY breaking masses -- see discussion below \Eref{vevs2}.

\begin{table}[!t]
\caption{\normalfont The mass spectrum of the model at the SUSY
vacuum}
\begin{tabular}{c@{\hspace{0.2cm}}|@{\hspace{0.2cm}}c@{\hspace{0.2cm}} |c}\toprule
\multicolumn{2}{c|}{Eigenstates}&\multicolumn{1}{c}{Eigenvalues}\\\cline{1-2}
Scalars &{Fermions} & (Masses)\\ \colrule
$\what P,\bar P$&${\lf\psi_{\bar P}\pm\psi_{\what P}\rg/\sqrt{2}}$ & {\hspace{0.2cm}} $\msn=m/\vev{J}$\\
$S,{\lf\delta\bX+\delta\Xb\rg/\sqrt{2}}$&${\lf\psi_{S}\pm\psi_+\rg/\sqrt{2}}$
& $m_{\rm PQ}=\sqrt{2}\la\Mpq$\\
${\lf\delta\bX-\delta\Xb\rg/\sqrt{2}}$&${\lf\psi_{\bX}-\psi_{\Xb}\rg/\sqrt{2}}$ & $0$\\
$\tilde\nu^c_i$&$\nu^c_i$ & $\mrh[i]=2\ld_{i\nu^c}\Mpq$\\
$D_{k{\rm a}},~\bar D_{k{\rm a}}$&$\psi_{D_{k{\rm a}}},~\psi_{\bar
D_{k{\rm a}}}$
& $m_{D_{\rm a}}=\ld_{D_{\rm a}}\Mpq$\\
$h_{l{\rm a}},~\bar h_{l{\rm a}}$&$\psi_{h_{l{\rm a}}},~\psi_{\bar
h_{l{\rm a}}}$ & $m_{h_{\rm a}}=\ld_{h_{\rm a}}\Mpq$\\ \botrule
\end{tabular}
\label{tab3}
\end{table}

The decay of $\what P$ commences when $\msn$ becomes larger than
the  expansion rate and is processed via the first coupling in the
RHS of Eq.~(\ref{Wnr}), into $S$ and $\ssni$ and $S, \ssni$ and
$\delta\Xb_{+}$ or $\delta\Xb_{-}$. The relevant Lagrangian sector
is
\beq \label{Ldc} {\cal L}_{\rm dc}=-{\msn\over\mP}{\ld_i}\what
P^*S\ssni \lf{\Mpq}+{\delta\Xb_{+}-\delta\Xb_{-}\over\sqrt{2}}\rg
+{\rm h.c.}\eeq
which arises from the cross term of the F-term, corresponding to
$\bar P$, of the SUSY potential derived from the superpotential
terms in \eqs{Whi}{Wnr}. Note that we have no $\ck$-induced decay
channels as in \cref{nmH}, since $\vev{P}=0$. The interaction
above gives rise to the following decay width
\beq \Gsn={1\over8\pi}\lf\lf{\Mpq\over
\mP}\rg^2+{1\over64\pi^2}\lf{\msn\over
\mP}\rg^2\rg\msn\sum_{i=1}^{3}\ld^2_i, \label{Gpq}\eeq
where we take into account that $\msn\gg\mpq$ and
$\msn\gg\mrh[j]$. These prerequisites are safely fulfilled when
$\la$ and $\ld_{i\nu^c}$ remain perturbative, i.e.
$\la,\ld_{i\nu^c}\leq\sqrt{4\pi}$ - see \Tref{tab3}. From the two
contributions to $\Gsn$, the dominant one is the second one -- the
3-body decay channel -- originating from the two last terms of
\Eref{Ldc}.

Taking also into account that the decay width of the produced
$\ssni$, $\Gamma_{i\nu^c}$, is much larger than \Gsn -- see below
-- we can infer that the reheat temperature, $\Trh$, is
exclusively determined by the $\what P$ decay and is given by
\cite{quin}
\beq \label{T1rh} \Trh=
\left(72\over5\pi^2g_{*}\right)^{1/4}\sqrt{\Gsn\mP},\eeq
where $g_{*}\simeq232.5$ counts the effective number of the
relativistic degrees of freedom at temperature $\Trh$ for the
(s)particle spectrum of MSSM plus the particle content of the
axion supermultiplet. Although the factor before the square root
of \Eref{T1rh} differs \cite{quin} slightly from other
calculations of $\Trh$ -- cf.~\cref{inlept} -- the numerical
result remains pretty stable and close to $10^8~\GeV$ -- see
\Sref{num1}.

If $\Trh\ll\mrh[i]$, the out-of-equilibrium condition \cite{baryo}
for the implementation of nTL is automatically satisfied.
Subsequently, $\tilde\nu^c_{i}$ decay into $\tilde H_u$ and $L_i$
or $\tilde H_u^*$ and $\tilde L_i^*$ via the tree-level couplings
derived from the second term of the second line of
Eq.~(\ref{wmssm}). Interference between tree-level and one-loop
diagrams generates a lepton-number asymmetry (per $\nu^c_{i}$
decay) $\ve_i$ \cite{baryo}, when CP is not conserved in the
Yukawa coupling constants $h_{Nij}$ -- see \Eref{wmssm}. The
resulting lepton-number asymmetry after reheating can be partially
converted through sphaleron effects into baryon-number asymmetry.
However, the required $\Trh$ must be compatible with constraints
for the $\Gr$ abundance, $Y_{\Gr}$, at the onset of
\emph{nucleosynthesis} (BBN). In particular, the $B$ yield can be
computed as
\beq Y_B=-0.35{5\over4}{\Trh\over\msn}\sum_i
\br_i\ve_i\>\>\>\mbox{with}\>\>\>\br_i={\ld^2_i\over\sum_i\ld_i^2}
\label{Yb}\eeq
the branching ratio of $\what P$ to $\ssni$ -- see \Eref{Gpq}. In
the formula above the first numerical factor ($0.35$) comes from
the sphaleron effects, whereas the second one ($5/4$) is due to
the slightly different calculation \cite{quin} of $\Trh$ --
cf.~\cref{inlept}. On the other hand, the $\Gr$ yield due to
thermal production at the onset of BBN is estimated to be
\cite{kohri}:
\beq\label{Ygr} Y_{\Gr}\simeq 1.9\cdot10^{-22} \Trh/\GeV.\eeq
where we assume that $\Gr$ is much heavier than the gauginos. Let
us note that non-thermal $\Gr$ production within SUGRA is unlikely
in our scenario, since these contributions are \cite{Idecay}
usually proportional to the v.e.v of the inflaton which is zero in
our case.

Both \eqs{Yb}{Ygr} calculate the correct values of the $B$ and
$\Gr$ abundances provided that no entropy production occurs for
$T<\Trh$ -- see also \Sref{cont1}. This fact can be easily
achieved within our setting. Indeed, following the arguments of
\cref{nmN}, one can show that the PQ system comprised of the
fields $S$ and $\delta\Xb_{+}$ decays via the third term in the
RHS of \Eref{Wnr} before its domination over radiation, for all
relevant values of $\ld_i$'s. Regarding the saxion,
$\delta\Xb_{-}$, we can assume that it has mass of the order of
$1~\TeV$, its decay mode to axions is suppressed (w.r.t the ones
to gluons, higgses and higgsinos \cite{Baer, senami, Covi}) and
the initial amplitude of its oscillations is equal to $f_a$. Under
these circumstances, it can \cite{Baer} decay before domination
too, and evades \cite{senami} the constraints from the effective
number of neutrinos for the $f_a$'s and $\Trh$'s encountered in
our model. As a consequence of its relatively large decay
temperature, the LSPs produced by the saxion decay are likely to
be thermalized and therefore, no upper bound on the saxion
abundance is \cite{senami} to be imposed. Finally, if axino is
sufficiently light it can act as a CDM candidate \cite{Baerax,
Covi} with relic abundance produced predominantly thermally -- due
to the relatively large $\Trh$. Otherwise, it may enhance
\cite{Covi} non-thermally the abundance of a higgsino-like
neutralino-LSP, rendering it a successful CDM candidate.

\subsection{\bf\scshape Lepton-Number Asymmetry and Neutrino Masses}\label{lept2}

As mentioned above, the decay of $\ssni$, emerging from the $\what
P$ decay, can generate a lepton asymmetry, $\ve_i$, caused by the
interference between the tree and one-loop decay diagrams,
provided that a CP-violation occurs in $h_{Nij}$'s. The produced
$\ve_i$ can be expressed in terms of the Dirac mass matrix of
$\nu_i$, $m_{\rm D}$, defined in a basis (called $\sni$-basis
henceforth) where $\sni$ are mass eigenstates, as follows:
\beqs\beq\ve_i ={\sum_{i\neq j}
\im\left[(\mD[]^\dag\mD[])_{ij}^2\right] \bigg( F_{\rm S}\lf
x_{ij},y_i,y_j\rg+F_{\rm V}(x_{ij})\bigg)
\over8\pi\vev{H_u}^2(\mD[]^\dag\mD[])_{ii}}, \label{el}\eeq where
we take $\vev{H_u}\simeq174~\GeV$, for large $\tan\beta$ and \beq
x_{ij}={\mrh[j]\over\mrh[i]}\>\>\>\mbox{and}\>\>\>y_i={\Gm[i\nu^c]\over\mrh[i]}=
{(\mD[]^\dag\mD[])_{ii}\over8\pi\vev{H_u}^2}\cdot\eeq Also
$F_{\rm V}$ and $F_{\rm S}$ represent, respectively, the
contributions from vertex and self-energy diagrams which in SUSY
theories read \cite{resonant1,resonant2,resonant3} \beq F_{\rm
V}\lf x\rg=-x\ln\lf1+ x^{-2}\rg,\eeq and \beq F_{\rm S}\lf
x,y,z\rg={-2x(x^2-1)\over\lf x^2-1-x^2z\ln x^2/\pi\rg^2+\lf
x^2z-y\rg^2},\eeq \eeqs with the latter expression written as
given in \cref{resonant3}. When \beq \Delta_{iji}\gg1\>\>\>
\mbox{and}\>\>\>\Delta_{ijj}\gg1\>\>\> \mbox{with}\>\>\>
\Delta_{ijk}={|x_{ij}^2-1|\over x_{ik} y_k},\label{Dijk}\eeq  (no
summation is  applied over the repeated indices) we can simplify
$F_{\rm S}$ expanding it close to $x\simeq1$ as follows \beq
F_{\rm S}\simeq{2x\over1-x^2}\simeq{1\over1-x}-{1\over2}\cdot\eeq
The involved in \Eref{el} $\mD[]$ can be diagonalized if we define
a basis -- called weak basis henceforth -- in which the lepton
Yukawa couplings and the $SU(2)_{\rm L}$ interactions are diagonal
in the space of generations. In particular we have
\beq \label{dD} U^\dag\mD[]U^{c\dag}=d_{\rm
D}=\diag\lf\mD[1],\mD[2], \mD[3]\rg,\eeq where $U$ and $U^c$ are
$3\times3$ unitary matrices which relate $L_i$ and $\sni$ (in the
$\sni$-basis) with the ones $L'_i$ and $\nu^{c\prime}_i$ in the
weak basis as follows:
\beq L'= L U\>\>\> \mbox{and}\>\>\>\nu^{c\prime}=U^c \nu^c.\eeq
Here, we write LH lepton superfields, i.e. $SU(2)_{\rm L}$ doublet
leptons, as row 3-vectors in family space and RH anti-lepton
superfields, i.e. $SU(2)_{\rm L}$ singlet anti-leptons, as column
3-vectors. Consequently, the combination $\mD[]^\dag\mD[]$
appeared in \Eref{el} turns out to be a function just of $d_{\rm
D}$ and $U^c$. Namely, \beq\mD[]^\dag\mD[]=U^{c\dag} d^\dag_{\rm
D}d_{\rm D}U^c. \label{mDD}\eeq

The connection of the leptogenesis scenario with the low energy
neutrino data can be achieved through the seesaw formula, which
gives the light-neutrino mass matrix $m_\nu$ in terms of $\mD[i]$
and $\mrh[i]$. Working in the $\sni$-basis, we have
\beq \label{seesaw} m_\nu= -m_{\rm D}\ d_{\nu^c}^{-1}\
m_{\rm D}^{\tr},\eeq where \beq\label{seesaw1} d_{\nu^c}=
\diag\lf\mrh[1],\mrh[2],\mrh[3]\rg \eeq with
$\mrh[1]\leq\mrh[2]\leq\mrh[3]$ real and positive.
Solving \Eref{dD} w.r.t $\mD[]$ and inserting the
resulting expression in \Eref{seesaw} we extract the mass matrix
\beq \label{bmn} \bar m_\nu=U^\dag m_\nu U^*=-d_{\rm
D}U^cd_{\nu^c}^{-1}U^{c\tr}d_{\rm D},\eeq which can be diagonalized
by the unitary PMNS matrix satisfying
\beq \bar m_\nu=U_\nu^*\ \diag\lf\mn[1],\mn[2],\mn[3]\rg\
U^\dag_\nu\label{mns1}\eeq and parameterized as follows:
\beq \label{mns2} U_\nu = \mtn{c_{12}c_{13}}{s_{12}c_{13}}{s_{13}
e^{-i\delta}} {U_{21\nu}}{U_{22\nu}}{s_{23}c_{13}}
{U_{31\nu}}{U_{32\nu}}{c_{23}c_{13}}\cdot {\cal P}. \eeq
Here \beqs\bea U_{21\nu}&=&-c_{23}s_{12}-s_{23}c_{12}s_{13} e^{i\delta},\\
U_{22\nu}&=& c_{23}c_{12}-s_{23}s_{12}s_{13} e^{i\delta},\\
U_{31\nu}&=& s_{23}s_{12}-c_{23}c_{12}s_{13} e^{i\delta},\\
U_{32\nu}&=&-s_{23}c_{12}-c_{23}s_{12}s_{13} e^{i\delta},\eea \eeqs
with $c_{ij}:=\cos \theta_{ij}$, $s_{ij}:=\sin \theta_{ij}$ and
$\delta$ the CP-violating Dirac phase. The two CP-violating
Majorana phases $\varphi_1$ and $\varphi_2$ are contained in the
matrix \beq {\cal P}=\diag\lf
e^{-i\varphi_1/2},e^{-i\varphi_2/2},1\rg.\eeq

Following a bottom-up approach, along the lines of \cref{senoguz},
we can find $\bar m_\nu$ via \Eref{mns1} using as input parameters
the low energy neutrino observables, the CP violating phases and
adopting the normal or inverted hierarchical scheme of neutrino
masses. Taking also $\mD[i]$ as input parameters we can construct
the complex symmetric matrix \beq W=-d_{\rm D}^{-1}\bar m_\nu
d_{\rm D}^{-1}=U^cd_{\nu^c}U^{c\tr}\label{Wm}\eeq -- see \Eref{bmn} --
from which we can extract $d_{\nu^c}$
as follows: \beq d_{\nu^c}^{-2}=U^{c\dag}W W^\dag
U^c.\label{WW}\eeq Note that $W W^\dag$ is a $3\times3$ complex,
hermitian matrix and can be diagonalized following the algorithm
described in \cref{33m}. Having determined the elements of $U^c$
and the $\mrh[i]$'s we can compute $\mD[]$ through \Eref{mDD} and
the $\ve_i$'s through \Eref{el}.

\section{Constraining the Model Parameters}\label{cont}

We exhibit the constraints that we impose on our cosmological
set-up in \Sref{cont1}, and delineate the allowed parameter space
of our model in Sec.~\ref{num}.

\subsection{\bf\scshape Imposed Constraints}\label{cont1}

The parameters of our model can be restricted once
we impose the following requirements:

\setcounter{paragraph}{0}

\paragraph{} According to the inflationary paradigm,
the horizon and flatness problems of the standard Big Bang
cosmology can be successfully resolved provided that
$\widehat{N}_{*}$ defined by \Eref{Nhi} takes a certain value,
which depends on the details of the cosmological scenario.
Employing standard methods \cite{nmi}, we can easily derive the
required $\widehat{N}_{*}$ for our model, consistent with the fact
that the PQ oscillatory system remains subdominant during the
post-inflationary era. Namely we obtain
\bea  \nonumber \widehat{N}_{*}&\simeq&22.5+2\ln{V_{\rm
CI}(\sg_{*})^{1/4}\over{1~{\rm GeV}}}-{4\over 3}\ln{V_{\rm C
I}(\sg_{\rm f})^{1/4}\over{1~{\rm GeV}}} \\ &+& {1\over3}\ln
{T_{\rm rh}\over{1~{\rm GeV}}}+{1\over2}\ln{f(\sg_{\rm f})\over
f(\sg_*)}\cdot \label{Ntot}\eea

\paragraph{} The inflationary observables derived in
\Sref{fhi2} are to be consistent with the fitting \cite{wmap} of
the WMAP7, BAO and $H_0$ data. As usual, we adopt the central
value of $\Delta_{\cal R}$, whereas we allow the remaining
quantities to vary within the 95$\%$ \emph{confidence level}
(c.l.) ranges. Namely,
\beqs\bea \label{obs1} \Delta_{\cal R}\simeq4.93\cdot 10^{-5},\\
0.944\leq \ns \leq0.992,\label{obs2}\\
-0.062\leq \as \leq0.018,\label{obs3} \\
r<0.24. \label{obs4}\eea\eeqs

\paragraph{} For the realization of nMCI, we assume that $\ck$ takes
relatively large values -- see e.g. \Eref{Vhi}. This assumption
may \cite{cutoff, unitarizing} jeopardize the validity of the
classical approximation, on which the analysis of the inflationary
behavior is based. To avoid this inconsistency -- which is rather
questionable \cite{cutoff, linde2} though -- we have to check the
hierarchy between the ultraviolet cut-off scale \cite{nmi},
$\Ld=\mP/\ck$, of the effective theory and the inflationary scale,
which is represented by $\Vhi(\sg_*)^{1/4}$ or, less
restrictively, by the corresponding Hubble parameter, $\widehat
H_*=\Vhi(\sg_*)^{1/2}/\sqrt{3}\mP$. In particular, the validity of
the effective theory implies \cite{cutoff}
\beq \label{Vl}\mbox{\ftn\sf (a)}\>\>\> \Vhi(\sg_*)^{1/4}\leq\Ld
\>\>\>\mbox{or}\>\>\>\mbox{\ftn\sf (b)}\>\>\> \widehat
H_*\leq\Ld.\eeq

\paragraph{} To ensure that the inflaton decay according to the
lagrangian part of \Eref{Ldc} is kinematically allowed we have to
impose the constraint -- see \Tref{tab3}:
\beq\label{kin}
\msn\geq2\mpq+\mrh[i]\>\>\>\Rightarrow\>\>\>2\mpq+\mrh[i]\lesssim10^{13}~\GeV,\eeq
where we make use of \Eref{mqa}. This requirement can be easily
satisfied by constraining $\la$ and $\ld_{i\nu^c}$ to values lower
than the perturbative limit. As the inequality in \Eref{kin} gets
strengthened, the accuracy of \Eref{Gpq} where masses of the
decay products are neglected, increases.

\paragraph{} From the solar, atmospheric, accelerator
and reactor neutrino experiments we take into account the
following inputs \cite{Expneutrino} -- see also
\cref{Lisi} -- on the neutrino mass-squared differences:
\beqs\bea \label{msol} \Delta m^2_{21}&=&\lf
7.59^{+0.2}_{-0.18}\rg\cdot10^{-3}~{\rm eV}^2,\\  \Delta
m^2_{31}&=&\lf
2.5^{+0.09}_{-0.16}\left[-2.4^{+0.08}_{-0.09}\right]\rg\cdot~10^{-3}~{\rm
eV}^2, \label{matm}\eea
on the mixing angles:
\bea\sin^2\theta_{12}&=&0.312^{+0.017}_{-0.015},\label{8atm1}\\
\sin^2\theta_{13}&=&0.013^{+0.007}_{-0.005}~\left[0.016^{+0.008}_{-0.006}\right],\label{8atm2}\\
\sin^2\theta_{23}&=&0.52^{+0.06}_{-0.07}~~~\left[0.52\pm0.06\right],
\label{8exp}\eea
and on the CP-violating Dirac phase:
\beq
\delta=-\lf0.61^{+0.75}_{-0.65}~\left[0.41^{+0.65}_{-0.7}\right]\rg\pi
\label{dexp}\eeq\eeqs for normal [inverted] neutrino mass
hierarchy. In particular, $\mn[i]$'s can be determined via the
relations\beqs \beq\mn[2]=\sqrt{\mn[1]^2+\Delta m^2_{21}}\eeq and
\beq\mn[3]=\sqrt{\mn[1]^2+\Delta m^2_{31}}\eeq for \emph{normally
ordered} (NO) $\mn[i]$'s or \beq\mn[1]=\sqrt{\mn[3]^2+\left|\Delta
m^2_{31}\right|}\eeq\eeqs for \emph{invertedly ordered} (IO)
$\mn[i]$'s. The sum of $\mn[i]$'s can be bounded from above by the
WMAP7 data \cite{wmap} \beq\label{sumn} \mbox{$\sum_i$}
\mn[i]\leq0.58~{\eV}\eeq at 95\% c.l. This is more restrictive
than the 95\% c.l. upper bound arising from the effective electron
neutrino mass in $\beta$-decay \cite{2beta}: \beq \label{mbeta}
m_{\beta}:=\left|\mbox{$\sum_i$}
U_{1i\nu}^2\mn[i]\right|\leq2.3~\eV.\eeq However, in the future,
the KATRIN experiment \cite{katrin} expects to reach the
sensitivity of $m_\beta\simeq0.2~\eV$ at $90\%$ c.l.

\paragraph{} The interpretation of BAU through nTL dictates \cite{wmap} at 95\% c.l.
%
%
\beq Y_B=\lf8.74\pm0.42\rg\cdot10^{-11}.\label{BAUwmap}\eeq

\paragraph{} In order to avoid spoiling the success of the
BBN, an upper bound on $Y_{\Gr}$ is to be imposed depending on the
$\Gr$ mass, $m_{\Gr}$,  and the dominant $\Gr$ decay mode. For the
conservative case where $\Gr$ decays with a tiny hadronic
branching ratio, we have \cite{kohri}
\beq  \label{Ygw} Y_{\Gr}\lesssim\left\{\bem
%
10^{-14}\hfill \cr
2.5\cdot10^{-14}\hfill \cr
4.3\cdot10^{-14}\hfill \cr
10^{-13}\hfill \cr\eem
\right.\>\>\>\mbox{for}\>\>\>m_{\Gr}\simeq\left\{\bem
0.69~{\rm TeV}\hfill \cr
5~{\rm TeV}\hfill \cr
8~{\rm TeV}\hfill \cr
10.6~{\rm TeV.}\hfill \cr\eem
\right.\eeq
As we see below, this bound is achievable  within our model model
only for $m_{\Gr}\gtrsim8~\TeV$. The bound above may be somehow
relaxed in the case of a stable $\Gr$.

\begin{figure*}[!t]
\centering
\includegraphics[width=60mm,angle=-90]{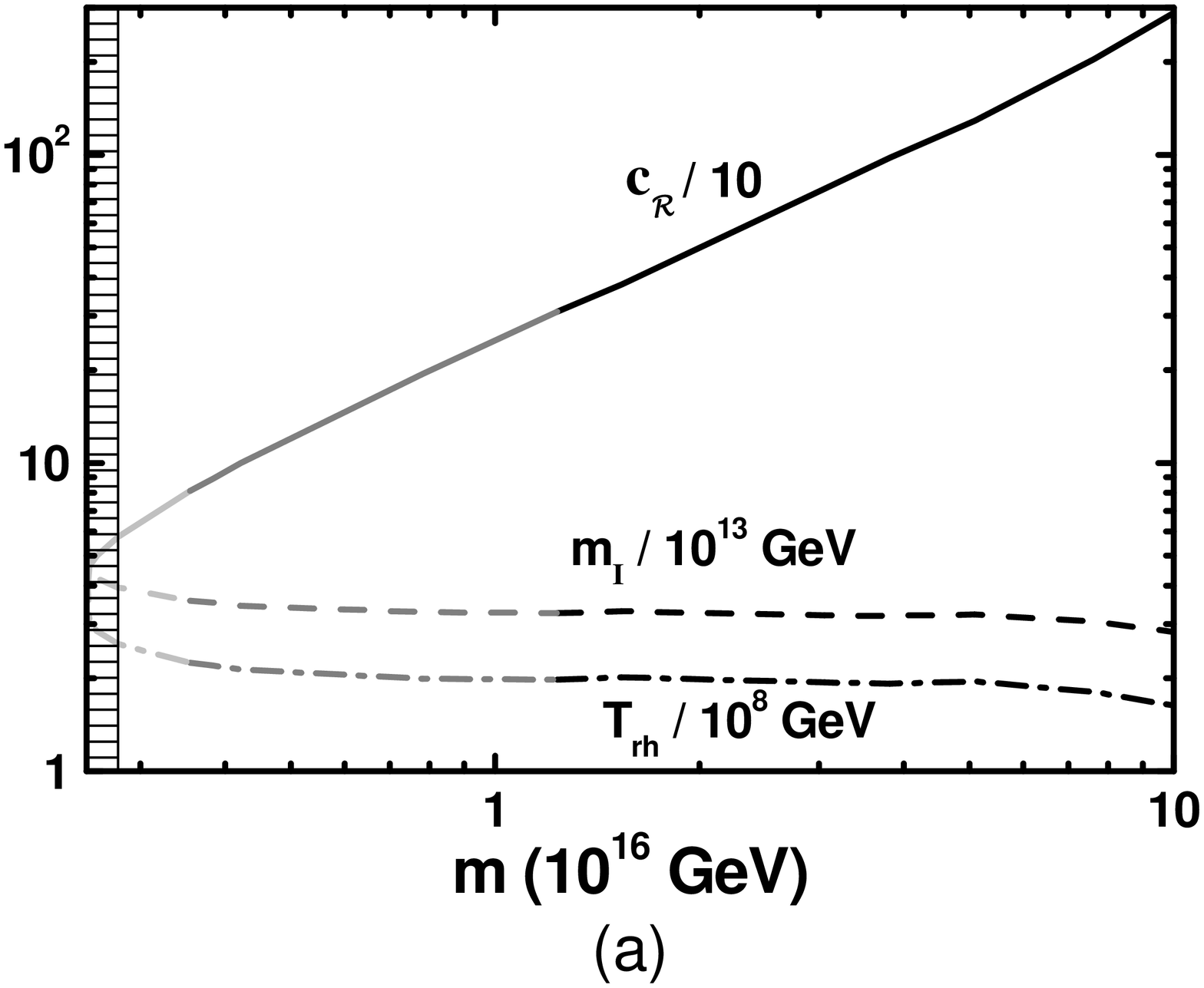}
\includegraphics[width=60mm,angle=-90]{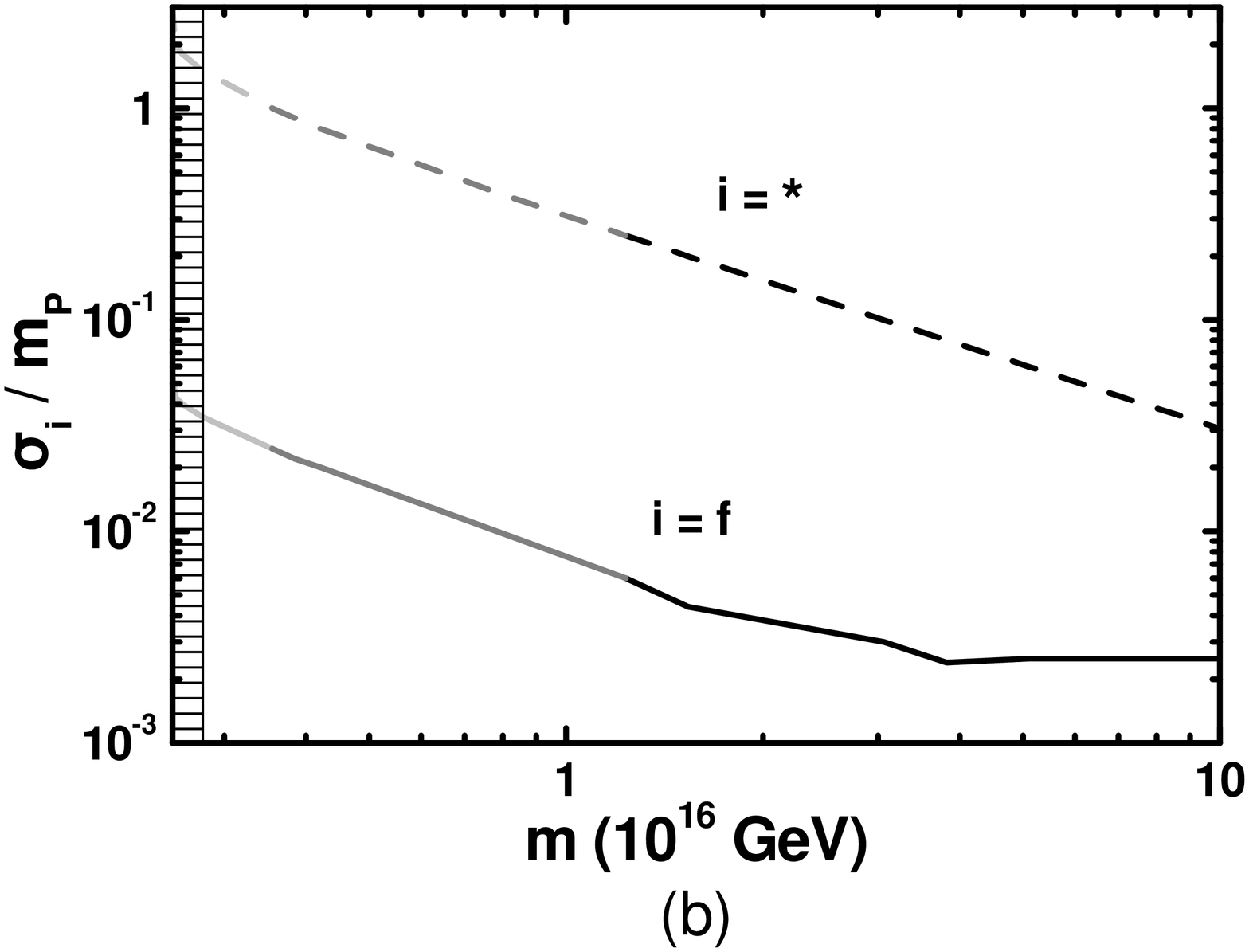}
\caption{\label{fig1}\sl The allowed by Eqs.~(\ref{Ntot}),
(\ref{obs1}), (\ref{obs2}) and (\ref{Vl}{\sf\small b}) values of
$\ck$ (solid line), $\msn$ -- given by \Eref{mqa} -- (dashed line)
and $\Trh$ -- given by \Eref{T1rh} -- (dot-dashed line) [$\sg_{\rm
f}$ (solid line) and $\sg_*$ (dashed line)] versus $m$ (a) [(b)]
for $\ld_1\ll\ld_2=\ld_3=0.5$. The light gray and gray segments
denote values of the various quantities satisfying \sEref{Vl}{a}
too, whereas along the light gray segments we obtain
$\sg_*\geq\mP$. Values of the parameters to the right of the lined
region correspond to $\ns$'s lying within its $68\%$ c.l.
observationally favored region.}
\end{figure*}

\subsection{\bf\scshape  Results}\label{num}

As can be easily seen from the relevant expressions in
Secs.~\ref{fhim} and \ref{lept2}, our cosmological set-up depends
on the following independent parameters:
$$m,~\la,~\lambda_\mu,~\kx,~\ld_i,~f_a,~{\rm n},
~\mn[\ell],~\mD[i],~\varphi_1~\mbox{and}~\varphi_2,$$ where
$\mn[\ell]$ is the low scale mass of the lightest of $\nu_i$'s and
can be identified with $\mn[1]~[\mn[3]]$ for NO [IO] neutrino mass
spectrum. We do not consider $\ck$ and $\ld_{i\nu^c}$ as
independent parameters since $\ck$ is related to $m$ via
\Eref{lan} while $\ld_{i\nu^c}$ can be derived from the last six
parameters above which affect exclusively the $Y_L$ calculation
and can be constrained through the requirements 5 and 6 of
\Sref{cont1}. Note that the $\ld_{i\nu^c}$'s can be replaced by
$\mrh[i]$'s given in \Tref{tab3} keeping in mind that
perturbativity requires $\ld_{i\nu^c}\leq\sqrt{4\pi}$ or
$\mrh[i]\leq 3.5f_a$. Recall also that $V_{\rm rc}$ in \Eref{Vrc}
is independent of $\ld_{D\rm a},\ld_{h\rm a}\gg\ld_a$ and depends
only on n, which is set equal to 5 for definiteness. To facilitate
the realization of see-saw mechanism, we take $f_a=10^{12}~\GeV$.
This choice makes also possible the generation of the $\mu$ term
of MSSM through the PQ symmetry breaking, since $\mu\sim1~\TeV$ is
obtained for $\lm=0.01$, whereas lower $f_a$'s dictate larger
$\lm$'s. Moreover, our computation reveals that $\ve_1$ in
\Eref{el} is mostly smaller than $\ve_2$ and $\ve_3$. Therefore,
fulfilling the baryogenesis criterion enforces us to consider
$\br_1\ll\br_{2,3}$ or $\ld_1\ll\ld_{2,3}\sim0.1$. Since $\ve_2$
and $\ve_3$ are of the same order of magnitude, the resulting
$Y_B$ does not depends crucially on $\ld_{2}/\ld_3$. Therefore we
believe that $\ld_{2}=\ld_3=0.5$ is a representative choice --
e.g., we explicitly checked that the option $\ld_{2}=0.1$ and
$\ld_3=0.9$ or $\ld_{2}=0.9$ and $\ld_3=0.1$ lead to similar
results. Finally, our results are independent of $\la$ and $\kx$
provided \Eref{kin} is fulfilled and the positivity of
$m_{\what{\bar p}}^2$ -- see \Tref{tab2} -- is ensured,
respectively. To facilitate the achievement of these objective, we
get $\la=0.01$ and $\kx=1$.

Summarizing, we set throughout our calculation:
\beqs\bea && \kx=1,~\ld_1\leq0.01,~\ld_2=\ld_3=0.5,~{\rm n}=5,\\
&& \ld_\mu=\la=0.01\>\>\mbox{and}\>\>f_a=10^{12}~\GeV.\eea\eeqs
The selected values for the above quantities give us a wide and
natural allowed region for the remaining fundamental parameters of
our model, as we show below concentrating separately in the
inflationary period (\Sref{num1}) and in the stage of nTL
(\Sref{num2}).

\subsubsection{\small \sf\scshape  The Stage of non-Minimal Inflation}\label{num1}

For nMCI, we use as input parameters in our numerical code
$\sigma_*, m$ and $\ck$. For every chosen $\ck\geq1$ we restrict
$m$ and $\sigma_*$ so that the conditions in \Eref{Ntot} -- with
$\Trh$ evaluated consistently using \Eref{T1rh} -- and
(\ref{obs1}) are satisfied. Let us remark that, in our numerical
calculations, we use the complete formulae for $\Vhi$ -- see
\Eref{Vhic} --, $\Ne_*$, the slow-roll parameters and
$\Delta_{\cal R}$ in Eqs.~(\ref{Nhi}), (\ref{sr1}), (\ref{sr2}),
(\ref{Prob}) and not the approximate relations listed in
\Sref{fhi2} for the sake of presentation.

Our results are displayed in \Fref{fig1}, where we draw the
allowed values of $\ck$ (solid line) $\msn$ (dashed line)and
$\Trh$ (dot-dashed line) [$\sg_{\rm f}$ (solid line) and $\sg_*$
(dashed line)] versus $m$ -- see \sFref{fig1}{a}
[\sFref{fig1}{b}]. The constraint of \sEref{Vl}{b} is satisfied
along the various curves whereas \sEref{Vl}{a} is valid only along
the gray and light gray segments of these. Along the light gray
segments, though, we obtain $\sg_*\geq\mP$. The lower bound on $m$
is derived from the saturation of the upper bound of inequality in
\Eref{obs2} whereas the upper bound comes from the fact that the
enhanced resulting $m$'s destabilize the inflationary path through
the radiative corrections in \Eref{Vhic} -- see \Eref{Vrc}.
Indeed, $V_{\rm rc}$ starts to influence the inflationary dynamics
for $m\geq1.5\cdot10^{16}~\GeV$, and consequently, the variation
of $\sg_{\rm f}$ as a function of $\ck$ or $m$ -- drawn in
\sFref{fig1}{b} -- deviates from the behavior described in
\Eref{sgap}. On the contrary the variations of $\sg_*$ follows
\Eref{s*}.

In all, we obtain
\beq\label{res1} 45\lesssim \ck\lesssim2950\>\>\>\mbox{and}\>\>\>
2.5\lesssim {m\over 10^{15}~\GeV}\lesssim102\eeq for
$\Ne_*\simeq54.5.$ From \sFref{fig1}{a}, we observe that $m$
depends on $\ck$ almost linearly whereas $\msn$ remains close to
$10^{13}~\GeV$ as we anticipated in \eqs{lan}{mqa}, respectively.
As a result of the latter effect, $\Trh$ given by \Eref{T1rh}
remains also almost constant. As $m$ (or $\ck$) decreases below
its maximal value in its allowed region in \Eref{res1}, we obtain
\beqs\bea\label{res} 0.965\lesssim \ns\lesssim 0.991, \\
6.5\lesssim {-\as/10^{-4}}\lesssim12  ,\\ 3.1\lesssim
{r/10^{-3}}\lesssim7.3.\eea\eeqs
Clearly, the predicted $\ns, \as$ and $r$ can lie within the
allowed ranges given in \eqss{obs2}{obs3}{obs4} respectively. In
particular, values of the various parameters plotted in
\Fref{fig1}, which lie to the right of the lined regions
correspond to $\ns\simeq(0.965-0.98)$. This result is consistent
with the $68\%$ c.l. observationally favored region -- see
\Eref{obs2}. It is notable, however, that $\ns$ increases
impressively for $\sg_*/\mP>\sqrt{6}$, contrary to the situation
in models of nMCI with quadratic coupling to $\rcc$ where $\ns$
remains constantly close to its central observational favored
value in \Eref{obs2} -- cf. \cref{nmN}.

As regards the $\Gr$ abundance, employing \Eref{Ygr}, we find \beq
3.5\lesssim {Y_{\Gr}/10^{-14}}\lesssim8.4\label{resgr}\eeq as $m$
varies within its allowed range in \Eref{res1}. Comparing  this
result with the limits of \Eref{Ygw}, we infer that our model is
consistent with the relevant restriction for
$m_{\Gr}\gtrsim(8-10)~\TeV$.

\subsubsection{\small \sf\scshape  The Stage of non-Thermal Leptogenesis}\label{num2}

As we show above, the stage of nMCI predicts almost constant
values of $\msn$ and $\Trh$ -- recall that we consider $\ld_i$'s
of the order of $0.1$. In other words, the post-inflationary
evolution in our set-up is largely independent of the precise
value of $m$ in the range of \Eref{res1}. As a consequence, $Y_B$
calculated by \Eref{Yb} does not vary with $m$, contrary to the
naive expectations. Just for definiteness we take throughout this
section $m=4.2\cdot10^{15}~\GeV$ which corresponds to
$\ck=100,~\ns=0.969,~\msn=3.4\cdot10^{13}~\GeV$ and
$\Trh=2.1\cdot10^{8}~\GeV$ ($Y_{\Gr}\simeq4\cdot10^{-14}$) --
recall that we use $\ld_1\ll\ld_2=\ld_3=0.5$.

On the contrary, $Y_B$ in our approach depends crucially on the
low energy parameters related to the neutrino physics. In our
numerical program, for a given neutrino mass scheme, we take as
input parameters: $\mn[\ell], \varphi_1, \varphi_2$ and the
best-fit values of the neutrino parameters listed in the paragraph
5 of \Sref{cont1}. We then find the \emph{renormalization group}
(RG) evolved values of these parameters at the scale of nTL,
$\Lambda_L$, which is taken to be $\Lambda_L=\msn$, integrating
numerically the complete expressions of the RG equations -- given
in \cref{running} -- for $\mn[i]$, $\theta_{ij}$, $\delta$,
$\varphi_1$ and $\varphi_2$. In doing this, we consider the MSSM
with $\tan\beta\simeq50$ (favored by the preliminary LHC results
\cite{mh2011}) as an effective theory between $\Lambda_L$ and a
SUSY-breaking scale, $M_{\rm SUSY}=1.5~\TeV$. Below $M_{\rm SUSY}$
the running of the various parameters is realized considering the
particle content of SM with a mass of about $120~\GeV$ for the
light Higgs. Following the procedure described in \Sref{lept2}, we
evaluate $\mrh[i]$ at $\Lambda_L$ taking $\mD[i]$ as free
parameters. In our approach we do not consider the running of
$\mD[i]$ and $\mrh[i]$  and therefore we give their values at
$\Lambda_L$.

\begin{table}[!t]
\caption{\normalfont Parameters yielding the correct BAU for
various neutrino mass schemes.}
\begin{tabular}{c|||c|c||c|c|c||c|c}\toprule
Parameters &  \multicolumn{7}{c}{Cases}\\\cline{2-8}
&A&B& C & D& E & F&G\\ \cline{2-8} &\multicolumn{2}{c||}{Normal} &
\multicolumn{3}{|c||}{Degenerate}&  \multicolumn{2}{|c}{Inverted}
\\& \multicolumn{2}{c||}{Hierarchy}&\multicolumn{3}{|c||}{Masses}& \multicolumn{2}{|c}{Hierarchy}\\
\colrule
\multicolumn{8}{c}{Low Scale Parameters}\\\colrule
$\mn[1]/0.1~\eV$&$0.05$&$0.1$&$0.5$ & $1.$& $0.7$ & $0.5$&$0.49$\\
$\mn[2]/0.1~\eV$&$0.1$&$0.13$&$0.51$ & $1.0$& $0.705$ & $0.51$&$0.5$\\
$\mn[3]/0.1~\eV$&$0.5$&$0.51$&$0.71$ & $1.12$&$0.5$ &
$0.1$&$0.05$\\\colrule
$\sum_i\mn[i]/0.1~\eV$&$0.65$&$0.74$&$1.7$ & $3.1$&$1.9$ &
$1.1$&$1$\\
$m_\beta/0.1~\eV$&$8\cdot10^{-3}$&$0.013$&$0.19$ & $0.46$&$0.3$ &
$0.42$&$0.44$\\\colrule
$\varphi_1$&$\pi$&$\pi$&$0$ & $\pi/4$&$\pi/4$ & $\pi/4$&$\pi/4$\\
$\varphi_2$&$0$&$0$ &$5\pi/6$& $\pi$&$\pi$ &
$\pi/2$&$\pi/4$\\\colrule
\multicolumn{8}{c}{Leptogenesis-Scale Parameters}\\\colrule
$\mD[1]/\GeV$&$2$&$2.5$&$4$ & $8$&$9$ & $6$&$5$\\
$\mD[2]/\GeV$&$3$&$3.49$&$5$ & $9.3$&$6$ & $3$&$1$\\
$\mD[3]/\GeV$&$6.7$&$4$&$8$ & $11$&$4.7$ & $2$&$2.1$\\\colrule
$\mrh[1]/10^{11}~\GeV$&$2.5$&$2.4$&$3.3$ & $6.5$&$4.6$ & $1$&$0.3$\\
$\mrh[2]/10^{11}~\GeV$&$11$&$7.3$&$5.2$ & $8.13$&$4.9$ & $5.56$&$4.3$\\
$\mrh[3]/10^{11}~\GeV$&$17$&$7.6$&$6$ & $8.36$&$8.6$ &
$6.7$&$5.1$\\\colrule
$\Delta_{iji}/10^4$&$4.5$&$0.5$&$0.96$ & $0.05$&$0.3$ & $1.6$&$1.4$\\
$\Delta_{ijj}/10^4$&$1.6$&$0.54$&$0.48$ & $0.04$&$0.3$ &
$1.2$&$3.5$\\\cline{2-8}
\multicolumn{8}{c}{(with $i=2$ and $j=3$ except for case E where
$i=1$ and $j=2$)}\\\colrule
\multicolumn{8}{c}{Resulting $B$-Yield }\\\colrule
$10^{11}Y^0_B$&$8.3$&$7.4$& $6.3$& $3.3$&$7.2$ & $9.3$&$4.8$\\
$10^{11}Y_B$&$8.7$&$8.85$& $8.98$& $8.4$&$8.9$ &
$8.96$&$8.95$\\\botrule
\end{tabular}
 \label{tab4}
\end{table}

\begin{figure*}[!t]
\centering
\includegraphics[width=60mm,angle=-90]{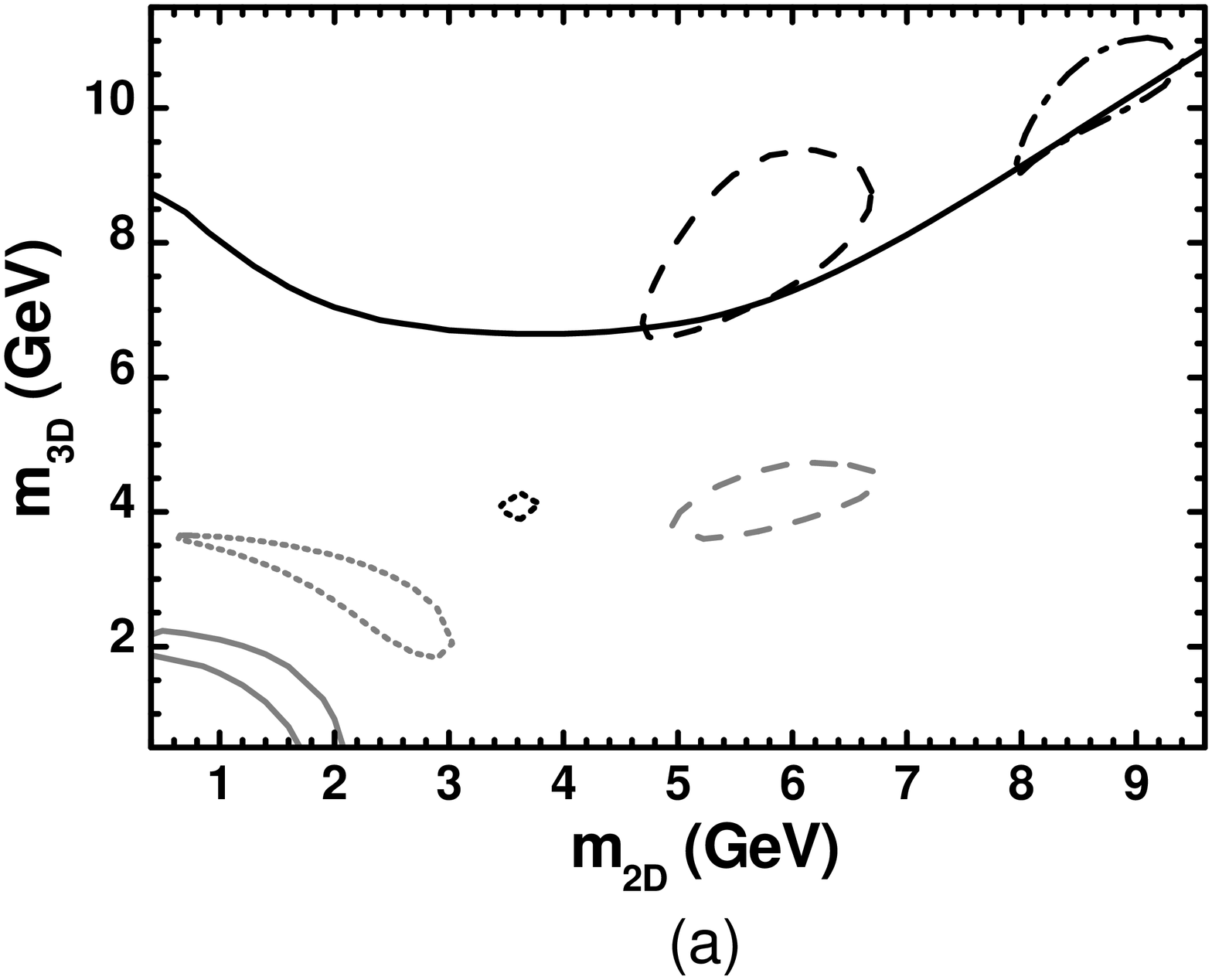}
\includegraphics[width=60mm,angle=-90]{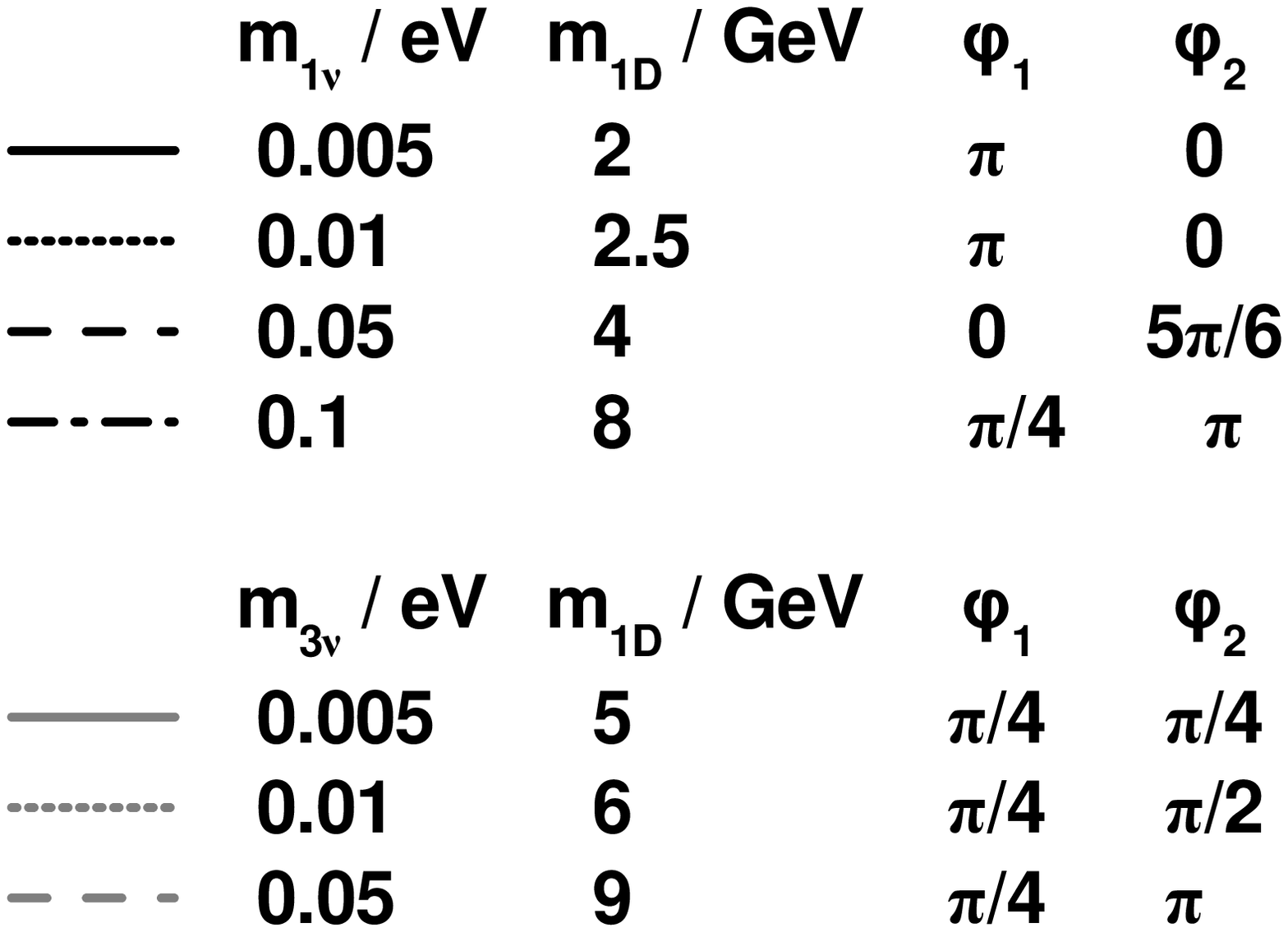}
\centering
\includegraphics[width=60mm,angle=-90]{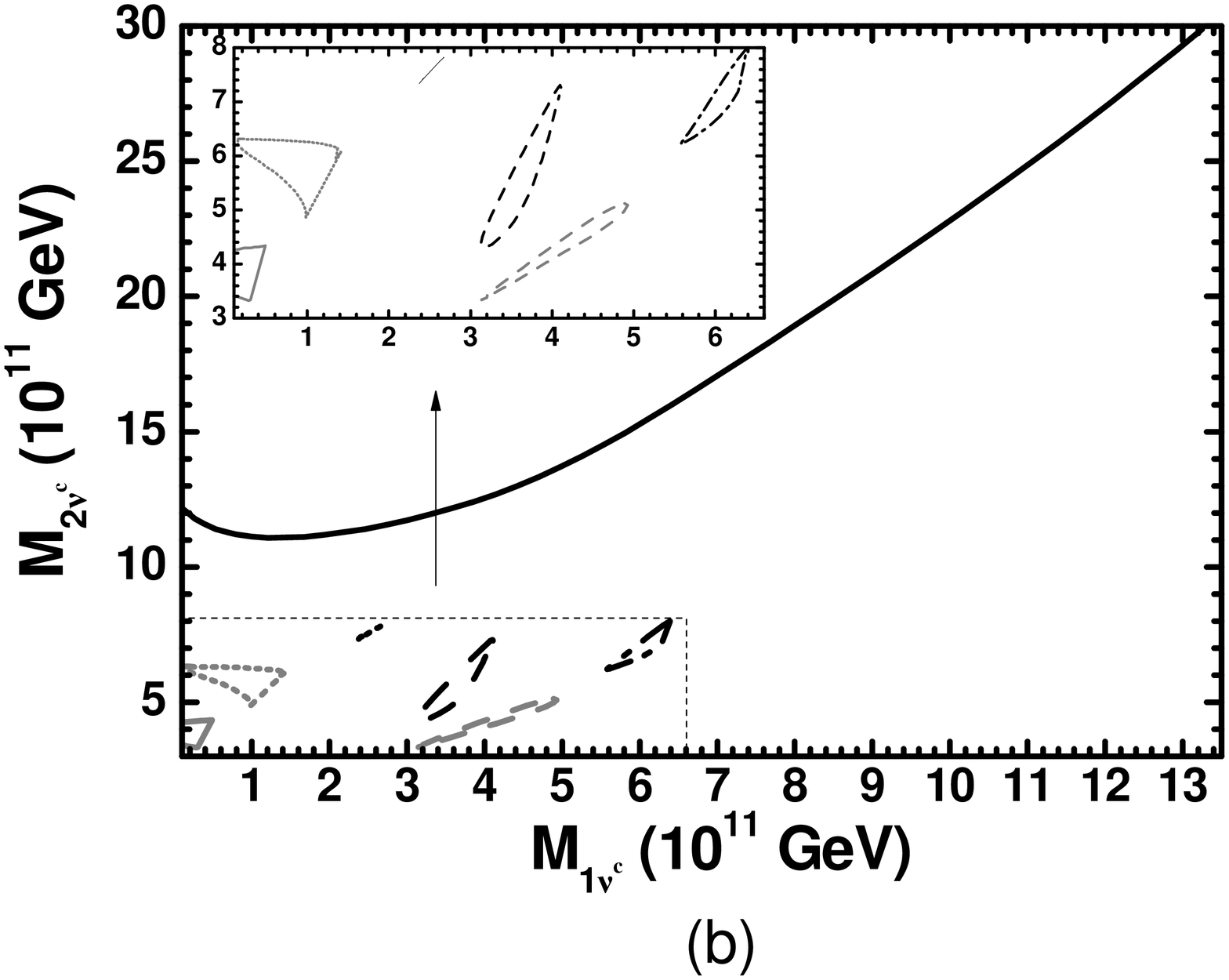}
\includegraphics[width=60mm,angle=-90]{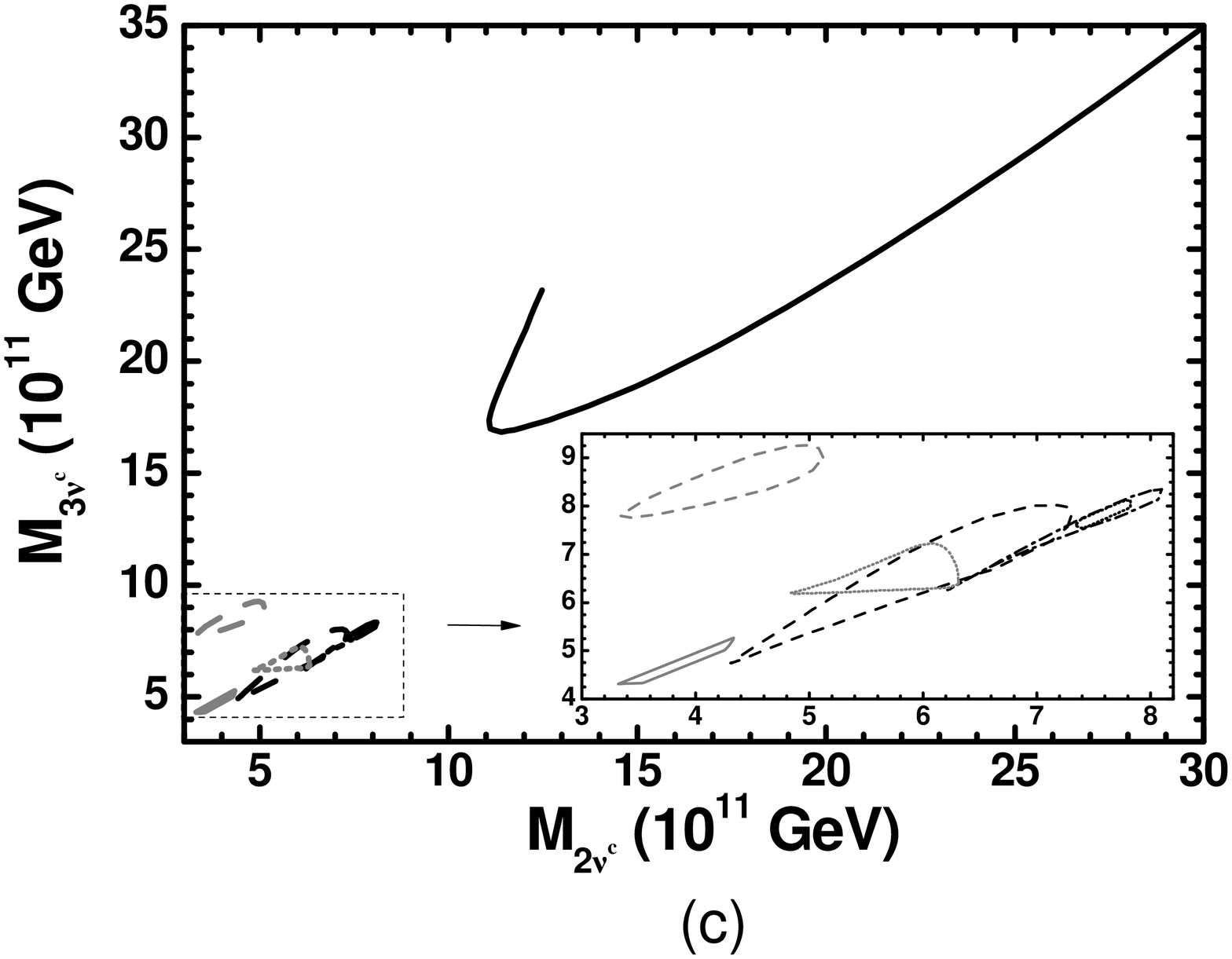}
\caption{\label{fig2}\sl Contours in the $\mD[2]-\mD[3]$ (a)
$\mrh[1]-\mrh[2]$ (b) and $\mrh[2]-\mrh[3]$ (c) plane yielding the
central $Y_B$ in \Eref{BAUwmap}, for various ($\mn[\ell], \mD[1],
\varphi_1, \varphi_2)$'s indicated next to the graph (a) and NO
[IO] $\mn[i]$'s (black [gray] lines).}
\end{figure*}

We start the exposition of our results arranging in \Tref{tab4}
some representative values of the parameters leading to the
correct BAU for normally hierarchical (cases A and B), degenerate
(cases C, D and E) and invertedly hierarchical (cases F and G)
neutrino masses. For comparison we display the $B$-yield with
($Y_B$) or without ($Y^0_B$) taking into account the RG effects.
We observe that the two results differ appreciably especially in
the cases with degenerate or IO $\mn[i]$'s. As it is evident from
the $\mD[i]$'s chosen, our model is not compatible with
any GUT-inspired pattern of large hierarchy between the
$\mD[i]$'s. In particular, we need $\mD[1]<\mD[2]<\mD[3]$
[$\mD[3]<\mD[2]<\mD[1]$] for NO [IO] $\mn[i]$'s (cases A, B, C and
D [cases E, F and G]).

From \Tref{tab4} we also notice that the achievement of $Y_B$
within the range of \Eref{BAUwmap} dictates mostly proximity
between two of the $\mrh[i]$'s. Indeed, except for the case A, we
obtain $\mrh[2]/\mrh[1]\simeq1.06$ in case E and
$\mrh[3]/\mrh[2]<1.2$ in the residual cases. However, it is clear
from the displayed $\Delta_{iji}$'s and $\Delta_{iji}$'s (with
$i=2$ and $j=3$ for all the cases besides case E where $i=1$ and
$j=2$) that in our framework the conditions of \Eref{Dijk} are
comfortably retained and therefore, our proposal is crucially
different from that of resonant leptogenesis
\cite{resonant1,resonant2,resonant3} -- it rather resembles that
of \cref{Asaka}. On the other hand, the correctness of $Y_B$ in
the case A entails $\mrh[2]$ and $\mrh[3]$ [$\ld_{2\nu^c}$ and
$\ld_{3\nu^c}$] roughly larger than $10^{12}~\GeV$ [unity]. In all
cases the current limit of \Eref{sumn} is safely met -- the case D
approaches it --, while $m_\beta$ turns out to be well below the
projected sensitivity of KATRIN \cite{katrin}.

To highlight further our conclusions inferred from \Tref{tab4}, we
can fix $\mn[\ell]$ ($\mn[1]$ for NO $\mn[i]$'s or $\mn[3]$ for IO
$\mn[i]$'s) $\mD[1]$, $\varphi_1$ and $\varphi_2$ to their values
shown in this table and vary $\mD[2]$ and $\mD[3]$ so that the
central value of \Eref{BAUwmap} is achieved. The resulting
contours in the $\mD[2]-\mD[3]$ plane are presented in
\sFref{fig2}{a} -- since the range of \Eref{BAUwmap} is very
narrow the possible variation of the drawn lines is negligible.
The resulting values of $\mrh[j]$ are displayed in
$\mrh[1]-\mrh[2]$ and $\mrh[2]-\mrh[3]$ plane -- see
\sFref{fig2}{b} and \sFref{fig2}{c} respectively. The conventions
adopted for the types and the color of the various lines are also
described next to the graphs (a) of \Fref{fig2}. In particular, we
use black [gray] lines for NO [IO] $\mn[i]$'s. Besides the case
with $\mn[1]=0.005~\eV$ we observe that every curve in all graphs
has two branches and not large hierarchies allowed in the sectors
of both the $\mD[]$'s and $\mrh[]$'s. Note that the black contour
for $\mn[1]=0.01~\eV$ in  \sFref{fig2}{c} is included within the
one for $\mn[1]=0.1~\eV$ and so, it is not quite distinguishable.
For $\mn[1]=0.005~\eV$, $\ld_{3\nu^c}$ saturates the perturbation
limit. Since we expect that the $\ld_{i\nu^c}$'s increase
\cite{suzuki} due to their RG running from low to higher scale,
our results do not jeopardize the validity of the conventional
perturbation approach up to the scale $\Lambda_L$. In all cases we
find $1\lesssim\mD[i]/\GeV\lesssim10$.

It is worth emphasizing that, although our mechanism of nTL is
connected with the specific inflationary model under
consideration, it can have a much wider applicability. It can be
realized within other models of inflation with similar inflaton
mass and reheat temperature, since it is largely independent of
the details of the  inflationary phase but restricts mainly the
yet unknown parameters of neutrino physics
($\mn[i],\mD[i],\varphi_1,\varphi_2$).

\section{Conclusions \label{con}}

We investigated a novel inflationary scenario in which the
inflaton field appears in a bilinear superpotential term and in a
linear holomorphic function included in a logarithmic \Ka. The
latter function can be interpreted in JF as a non-minimal coupling
to gravity, whose the strength is constrained so as the EF
inflationary potential can be flattened enough to support a stage
of non-minimal inflation compatible with observations. The
inflationary model was embedded in a moderate extension of MSSM
augmented by three RH neutrino superfields and three other singlet
superfields, which lead to a PQPT tied to renormalizable
superpotential terms. The PQPT follows \FHI and resolves the
strong CP and the $\mu$ problems of MSSM and also provides RH
neutrinos with masses lower than about $10^{12}~\GeV$. The
possible catastrophic production of domain walls can be eluded by
the introduction of extra matter superfields which can be chosen
so that the MSSM gauge coupling constant unification is not
disturbed. For $\Gr$ masses larger than $8~{\rm TeV}$,
observationally safe reheating of the universe with
$\Trh\simeq10^8~\GeV$ can be accomplished by a three-body decay of
the inflaton. The subsequent out-of-equilibrium decays of the
produced RH sneutrinos can generate the required by the
observations BAU consistently with the present low energy neutrino
data, provided that the Dirac neutrino masses are constrained in
the range $(1-10)~\GeV$ for all the light neutrino mass schemes.
It is gratifying that the degeneracy of the masses of the RH
(s)neutrinos required by the mechanism of nTL in our model is low
enough compared with their decay widths, so that perturbative
calculation remains safely valid. Finally, we briefly discussed
scenaria in which the potential axino and saxion overproduction
problems can be avoided.

\acknowledgments C.P. acknowledges the Bartol Research Institute
and the Department of Physics and Astronomy of the University of
Delaware for its warm hospitality, during which this work has been
initiated. Q.S. acknowledges support by the DOE grant No.
DE-FG02-12ER41808. We would like to thank G.~Lazarides, H.M.~Lee,
W.-I.~Park, M.Ur Rehman and N.~Toumbas for helpful discussions.


\def\ijmp#1#2#3{{\sl Int. Jour. Mod. Phys.}
{\bf #1},~#3~(#2)}
\def\plb#1#2#3{{\sl Phys. Lett. B }{\bf #1}, #3 (#2)}
\def\prl#1#2#3{{\sl Phys. Rev. Lett.}
{\bf #1},~#3~(#2)}
\def\rmp#1#2#3{{Rev. Mod. Phys.}
{\bf #1},~#3~(#2)}
\def\prep#1#2#3{{\sl Phys. Rep. }{\bf #1}, #3 (#2)}
\def\prd#1#2#3{{\sl Phys. Rev. D }{\bf #1}, #3 (#2)}
\def\npb#1#2#3{{\sl Nucl. Phys. }{\bf B#1}, #3 (#2)}
\def\npps#1#2#3{{Nucl. Phys. B (Proc. Sup.)}
{\bf #1},~#3~(#2)}
\def\mpl#1#2#3{{Mod. Phys. Lett.}
{\bf #1},~#3~(#2)}
\def\jetp#1#2#3{{JETP Lett. }{\bf #1}, #3 (#2)}
\def\app#1#2#3{{Acta Phys. Polon.}
{\bf #1},~#3~(#2)}
\def\ptp#1#2#3{{Prog. Theor. Phys.}
{\bf #1},~#3~(#2)}
\def\n#1#2#3{{Nature }{\bf #1},~#3~(#2)}
\def\apj#1#2#3{{Astrophys. J.}
{\bf #1},~#3~(#2)}
\def\mnras#1#2#3{{MNRAS }{\bf #1},~#3~(#2)}
\def\grg#1#2#3{{Gen. Rel. Grav.}
{\bf #1},~#3~(#2)}
\def\s#1#2#3{{Science }{\bf #1},~#3~(#2)}
\def\ibid#1#2#3{{\it ibid. }{\bf #1},~#3~(#2)}
\def\cpc#1#2#3{{Comput. Phys. Commun.}
{\bf #1},~#3~(#2)}
\def\astp#1#2#3{{Astropart. Phys.}
{\bf #1},~#3~(#2)}
\def\epjc#1#2#3{{Eur. Phys. J. C}
{\bf #1},~#3~(#2)}
\def\jhep#1#2#3{{\sl J. High Energy Phys.}
{\bf #1}, #3 (#2)}
\newcommand\jcap[3]{{\sl J.\ Cosmol.\ Astropart.\ Phys.\ }{\bf #1}, #3 (#2)}
\newcommand\njp[3]{{\sl New.\ J.\ Phys.\ }{\bf #1}, #3 (#2)}

\end{document}